\documentclass[journal]{IEEEtran}

\usepackage{xcolor,soul,framed} 

\colorlet{shadecolor}{yellow}
\usepackage[pdftex]{graphicx}
\graphicspath{{../pdf/}{../jpeg/}}
\DeclareGraphicsExtensions{.pdf,.jpeg,.png}

\usepackage[cmex10]{amsmath}
\usepackage{array}
\usepackage{mdwmath}
\usepackage{mdwtab}
\usepackage{eqparbox}
\usepackage{url}
\usepackage{hyperref}
\usepackage{breakurl}
\usepackage{subfigure}

\begin{document}
\bstctlcite{IEEEexample:BSTcontrol}
    \title{Reflectance of Silicon Photomultipliers at Vacuum Ultraviolet Wavelengths}
    \author{P.~Lv, G.F.~Cao, L.J.~Wen, S.~Al Kharusi, G.~Anton, I.J.~Arnquist, I.~Badhrees, P.S.~Barbeau, D.~Beck, V.~Belov
% ...
    

  
  \thanks{Please see the Acknowledgment section of this paper for the author affiliations.}}%


\maketitle

\begin{abstract}
Characterization of the vacuum ultraviolet (VUV) reflectance of silicon photomultipliers (SiPMs) is important for large-scale SiPM-based photodetector systems. We report the angular dependence of the specular reflectance in a vacuum of SiPMs manufactured by Fondazionc Bruno Kessler (FBK) and Hamamatsu Photonics K.K. (HPK) over wavelengths ranging from 120 nm to 280 nm. Refractive index and extinction coefficient of the thin silicon-dioxide film deposited on the surface of the FBK SiPMs are derived from reflectance data of a FBK silicon wafer with the same deposited oxide film as SiPMs. The diffuse reflectance of SiPMs is also measured at 193~nm. We use the VUV spectral dependence of the optical constants to predict the reflectance of the FBK silicon wafer and FBK SiPMs in liquid xenon.
\end{abstract}


%
\IEEEpeerreviewmaketitle


\section{Introduction}

\IEEEPARstart{T}{he} SiPM is a novel solid-state silicon photon detector, connecting arrays of avalanche photon diodes (APDs) in parallel on a common silicon substrate \cite{sipm}. Each APD is operated in Geiger mode and coupled with a quenching resistor. In recent decades, the VUV performance of SiPMs has been significantly improved, with reduced cost that warrants affordable meter-square-scale arrays of a SiPM photodetector system. Compared to traditional photomultiplier tubes (PMTs), SiPMs are more compact, have high radio purity, and exhibit good photon detection efficiency (PDE). These features make SiPMs more attractive for applications in cryogenic experiments, in particular in rare-event searches \cite{nexo_cdr} \cite{dune} \cite{meg2} \cite{darkside} \cite{darwin}. There SiPMs benefit from applications in cryogenic environments that reduce the dark noise rate to 5~Hz/mm$^2$ at liquid xenon temperatures compared to rates of 50-100~kHz/mm$^2$ at room temperature. The absolute PDE is a key performance parameter of SiPMs. It is related to the fill factor, transmittance of SiPM surface layers, quantum efficiency (QE), and avalanche trigger probability. One way to improve the PDE is to design antireflective coatings (ARCs) on the SiPM surface to enhance the photon transmittance. In contrast to the visible region, this enhancement is essential for VUV wavelengths, since the real part of the refractive index of silicon is less than one \cite{si_nk} in the VUV region and it is much smaller than that of intrinsic silicon dioxide (SiO$_{2}$) layer or other suitable ARC materials. Over 50\% of VUV photons can be reflected by the SiPM surface with a single layer of thin SiO$_{2}$, due to the large index mismatch. However, in a large photodetector system, reflected photons can be captured by other SiPMs and contributed to the overall light collection efficiency. Knowledge on the VUV reflectance of SiPMs will allow us to better understand the optical response and the performance of SiPM photodetector systems. It has become an important R\&D topic in large-scale photodetector programs, such as the nEXO. The absolute PDE of SiPMs at VUV wavelengths is investigated in Ref.\cite{nexo_pde1} \cite{nexo_pde2} \cite{nexo_pde3} \cite{meg_pde}; however, little is known regarding the VUV reflectance of SiPMs.

The nEXO experiment is being designed to search for 0$\nu\beta\beta$ decays in 5 tonnes of liquid xenon (LXe) enriched in the isotope $^{136}$Xe in a time-projection chamber (TPC). nEXO's projected sensitivity is close to 10$^{28}$ years after 10 years of data taking \cite{nexo_sen}. Instead of the large-area APD used in detectors such as EXO-200 \cite{exo-200}, a 4-5 m$^{2}$ SiPM array is proposed for the detection of the $\sim$175~nm scintillation photons from LXe \cite{nexo_cdr}. In combination with the information on charge detection in the TPC, the anticipated energy resolution is projected to be 1\% at Q$_{\beta\beta}$. However, the overall photon collection efficiency of the photodetector system is one of the major factors that will impact the energy resolution. The overall photon collection efficiency can be further classified into the photon transport efficiency (PTE) and PDE of SiPMs. The PTE can be quantified by a full Monte Carlo simulation with detailed geometry implementations and the knowledge of the optical properties of components inside the TPC. The VUV reflectance of SiPMs in LXe is one of the input parameters in such simulations conducted to accurately predict the PTE. 

The nEXO collaboration has built a dedicated optical setup to study the reflectance of SiPMs in LXe, where SiPMs from HPK have been measured recently \cite{lixo}. However, a vacuum-based setup has a wider VUV spectral range, and the reflectance in a vacuum (or argon- or nitrogen-purged setups) can guide us in verifying the results of LXe measurements. Establishing a predictable relationship between vacuum and LXe environments would be much more efficient and convenient than performing additional measurements in LXe, which are costly and complex.

This manuscript is organized as follows. First, we discuss the instrumentation and sample characteristics. We quantify the measurement uncertainties. Then, we present the measurement results of VUV specular and diffuse reflectance. We derive the optical constants and the thickness of the SiO$_{2}$ film intrinsically deposited on the FBK SiPM surface. Finally, we use the VUV spectral dependence of the optical constants to predict the reflectance of the FBK silicon wafer and SiPMs in LXe.

\section{Instrumentation}

\subsection{The specular reflection optical system}
\begin{figure}
\centering
\includegraphics[width=3.4in]{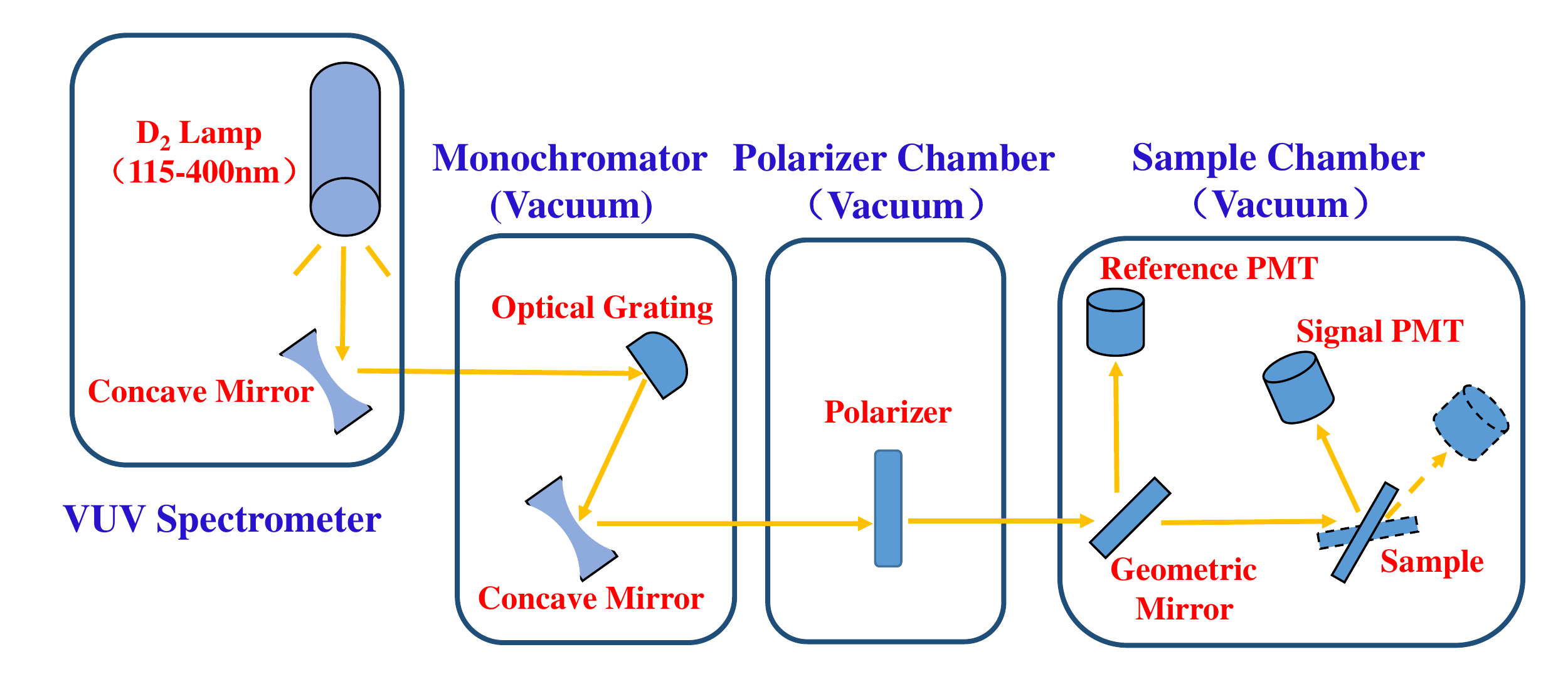}
\caption{Schematic diagram of the specular reflection optical system. (From left to right: VUV spectrometer, monochromator chamber, polarization chamber and sample chamber.)} \label{Specular_system}
\end{figure}

The VUV spectrophotometer of the specular reflection optical system is provided by Laser Zentrum Hannover, Germany, and its schematic diagram is presented in Figure \ref{Specular_system}. Two deuterium lamps with a magnesium fluoride (MgF$_2$) window and a quartz window serve as illuminants, which emit photons with wavelengths from 115~nm to 230~nm and from 170~nm to 300~nm, respectively. The spectral range can be selected during the measurement as long as the vacuum seal is not broken. The light beam is focused with a concave mirror onto the entry slit of a monochromator. The monochromator chamber contains an optical grating system used to select the wavelength and a second concave mirror to direct the irradiation into the sample chamber. The widths of the entrance slit and the exit slit are set at 100~$\mu$m, corresponding to a wavelength resolution of 0.8~nm. A polarization chamber can be inserted into the region between the monochromator chamber and the sample chamber to select incident light with a specific polarization. The sample chamber consists of a geometric mirror, a signal PMT with a UV-converter coating to detect reflected light, a reference PMT to monitor the stability of the light beam intensity and provide the intensity of incident light on samples based on the light intensity ratio between the signal PMT and the reference PMT, and movable units to rotate samples and the signal PMT. The incident angle onto the sample can be varied from 8 degree to 55 degree and it is automatically adjusted by software. The sample chamber is connected to a molecular pump, which can provide a $10^{-1}$~mbar vacuum for the whole system. The profile of the light beam is measured to be 3~mm x 5~mm, with the shape of a rectangle. By moving the sample away from the light beam, the signal PMT measures the intensity of the incident lights, and the light intensity ratio between the signal PMT and the reference PMT is obtained at different wavelengths. With a fixed angle of incidence, the signal PMT is rotated to search for the maximum intensity of reflected light; then, the specular reflectance is calculated by using the ratio of the intensities of the reflected and incident light.  An aperture is placed in front of the signal PMT with a set diameter of 4~mm. The distance from the sample to the signal PMT is 90~mm; Subsequently, the acceptance angle of the reflected light is calculated to be 1.55$\times$10$^{-3}$ radian.

\subsection{The total integrated scatter setup}
\begin{figure}
\centering
\includegraphics[width=3.4in]{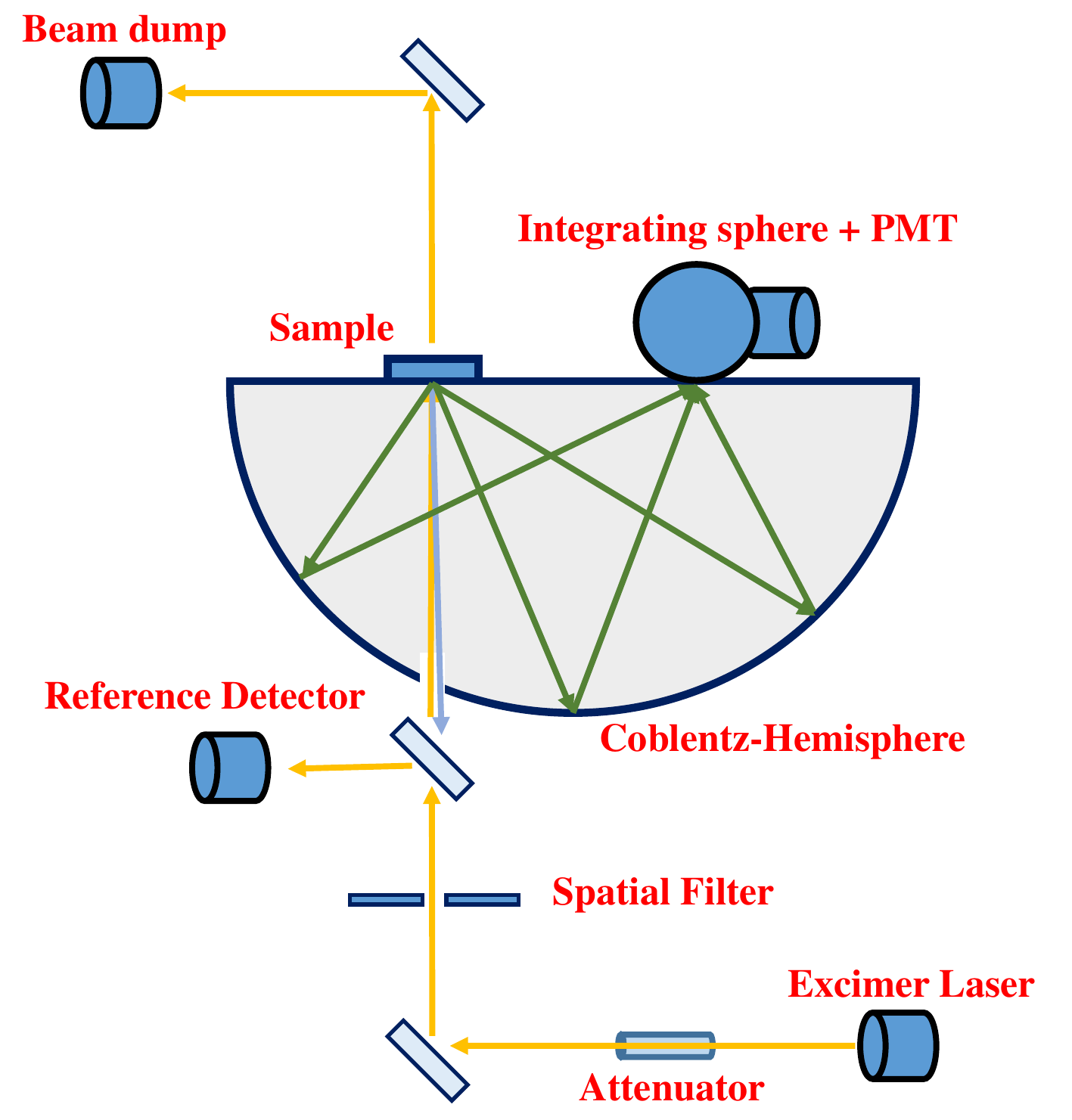}
\caption{Diagram of the total integrated scatter setup. (The yellow, blue and green lines represent incident light, specular reflection light and scattered light, respectively.)} 
\label{Diffuse_system}
\end{figure}

The setup for the total integrated scatter (TIS) is manufactured by Laser Zentrum Hannover, the Department of Laser Components. The diagram of the setup is shown in Figure \ref{Diffuse_system}, which is based on a Coblentz hemisphere with a diameter of 350~mm, designed to measure the low level of scatterings from the sample surfaces, even from optically smooth surfaces. The inner surface of the hemisphere is coated with an aluminum film and a protection layer to prevent aluminum oxidation. The aluminum film serves as a mirror and focuses light onto the detector. The light beam generated by a pulsed 193~nm laser is attenuated and guided onto the sample through the incident port, which has a diameter of 10~mm and an angle of incidence close to 0 degree. Specular reflected light from the sample leave from the same port of incident light which has an open angle of 2$\times10^{-4}$ radian. Scattered light is collected by an integrating sphere with a UV-converter coating and then detected by a PMT attached to the sphere. The size of the light beam is 100~$\mu$m at the sample position, and the intensity of the light beam is monitored by a reference PMT via applying a beam splitter to the light beam. The sample to be measured sits on a 2D transportation platform, which is used to scan the surface of the sample with a scan length of 50~mm in two directions. The resolution and repeatability of positioning are better than 100~$\mu$m and 200~$\mu$m, respectively. The detector and the sample to be measured are positioned at conjugate foci of the Coblentz hemisphere.

\section{The measured samples}
In this work, six samples are measured by a specular reflection optical system, and five of them are measured by the TIS. They are summarized in Table {\ref{tab:1}}. Four samples are provided by FBK. FBK-VUV-HD1-LF and FBK-VUV-HD1-STD are two types of VUV sensitive SiPMs developed in 2017 by FBK. These two FBK SiPMs have the same dimensions (10~mm x 10~mm), and the pixel size is 30~$\mu$m, which yields a filling factor of $\sim$73\%. To eliminate the influence on reflectance from the microstructure on the surface of SiPMs, FBK manufactured a six-inch silicon wafer deposited by a layer of SiO$_2$ with a thickness of approximately 1.5~$\mu$m. This silicon wafer is identical to the one used during SiPM manufacturing and diced into 20~mm X 20~mm pieces. Two of the pieces are selected to measure reflectance in this paper. The remaining two samples are provided by HPK, which are the fourth generation of VUV-sensitive SiPMs (Hamamatsu-VUV4) with dimensions of 6~mm x 6~mm. The series numbers are S13370-6050CN and S13370-6075CN, corresponding to pixel sizes of 50~$\mu$m and 75~$\mu$m, respectively. The VUV4 SiPM with a larger pixel size has a larger filling factor. 

\begin{table}
	\centering
	\caption{\label{tab:1} List of measured samples.}
	\scalebox{0.9}{
	\smallskip
	\begin{tabular}{|c<{\centering}|p{15mm}<{\centering}|p{12mm}<{\centering}|c|}
		\hline
		Sample Name& Dimensions (mm $\times$ mm)& Pixel Size ($\mu$m)& Filling Factor\\
		\hline	
		FBK-Si-Wafer \#1    & 20 $\times$ 20&-- &-- \\
		FBK-Si-Wafer \#2    & 20 $\times$ 20&-- &-- \\
		FBK-VUV-HD1-LF      &10 $\times$ 10&~30 &73\% \\
		FBK-VUV-HD1-STD    &10 $\times$ 10&~30 &73\%	\\
		Hamamatsu-VUV4 \#1 &6 $\times$ 6&~50	  &60\%	\\
		Hamamatsu-VUV4 \#2 &6 $\times$ 6&~75    &70\%	\\
		\hline
	\end{tabular}}
\end{table}

\section{Estimation of measurement uncertainties}
\begin{figure}
\centering
\includegraphics[width=3in]{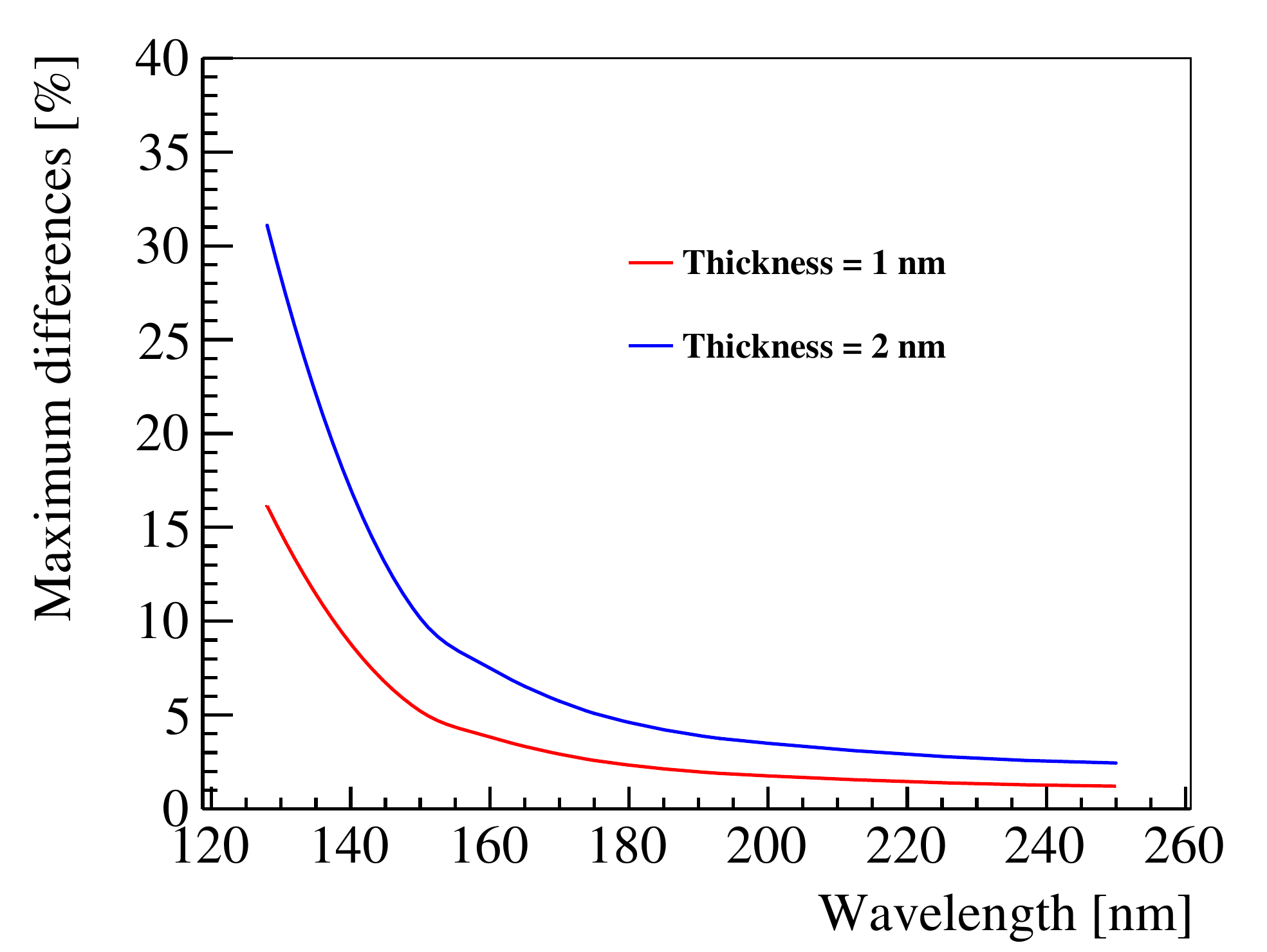}
\caption{Maximum differences in calculated reflectance over the full range of incident angles with thicknesses of 1~nm (red) and 2~nm (blue) of the oxide layer compared to that without the oxide layer as a function of wavelengths.} 
\label{maxdiff}
\end{figure}

\begin{figure}
    \centering
    \subfigure{\includegraphics[width=3in]{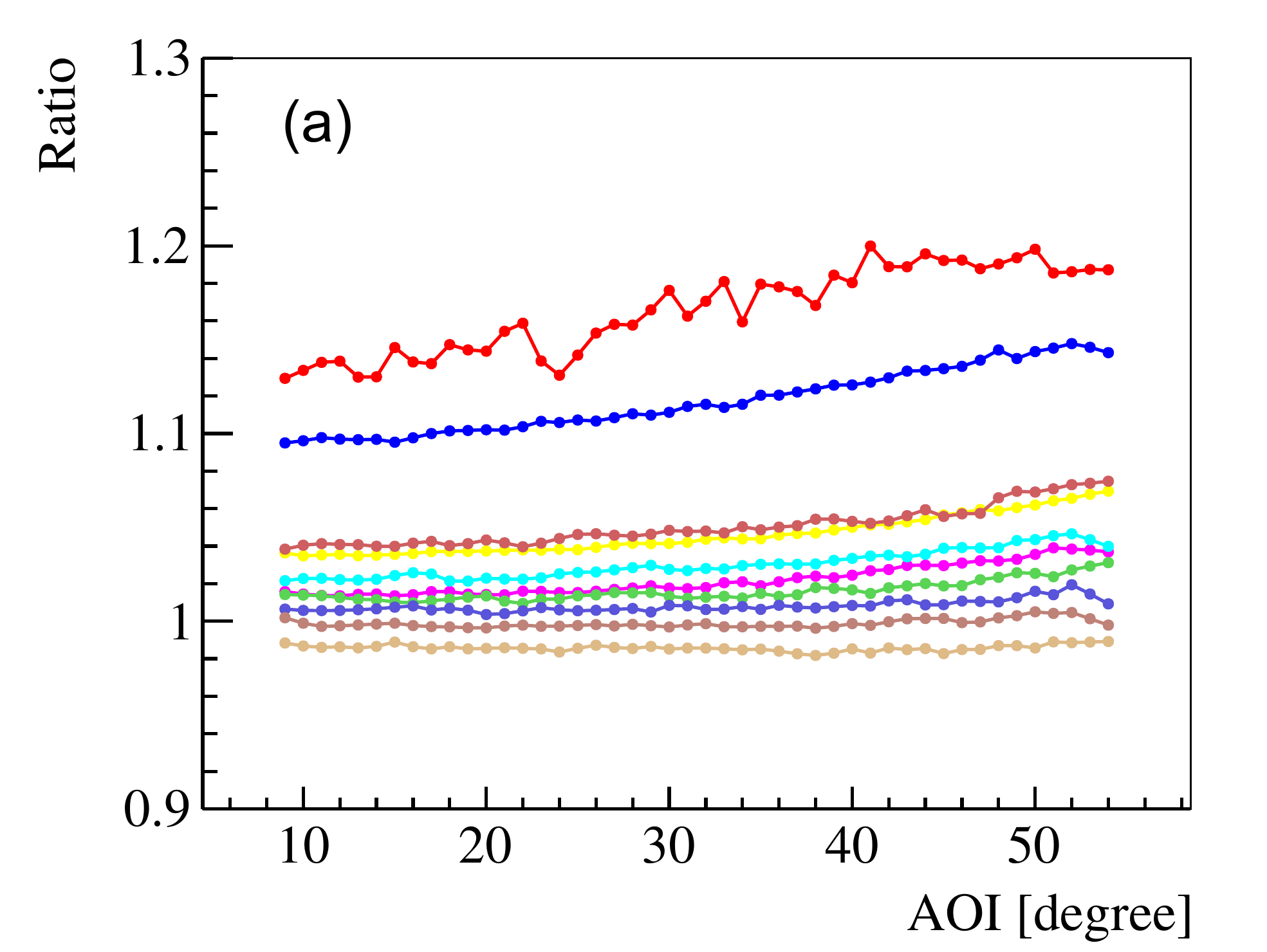}}
	\subfigure{\includegraphics[width=3in]{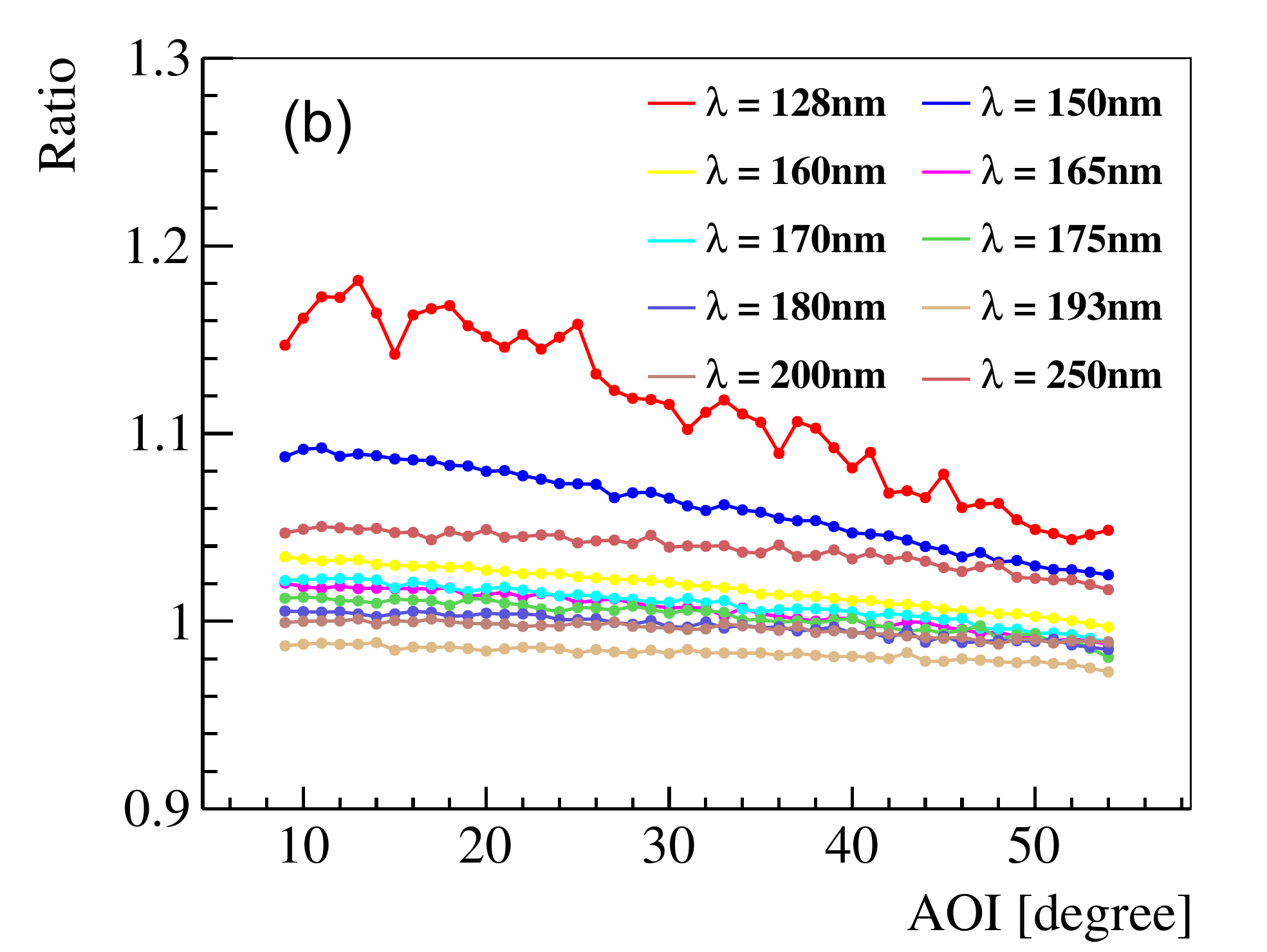}}
    \caption{Ratios of the measured reflectance and the calculated reflectance for the reference sample (assuming no oxide layer on the sample) as a function of angle of incidence (AOI) at ten different wavelengths. (a) indicates s-polarization, and (b) is p-polarization.}
    \label{ratio}
\end{figure}

A pure silicon wafer is used as a reference sample to estimate uncertainties induced by the specular reflectance optical system. A native oxide layer exists on the surface of the reference sample, since the wafer has been exposed to air for a long time (more than 1 year). Thus, before the reference sample is delivered to the sample chamber, it undergoes three chemical cleaning processes, as discussed in Ref. \cite{wafer_cleaning}, to remove the organic residuals, metal contamination and native oxide layer on the wafer surface. However, after cleaning, a thin native oxide layer is still expected on the wafer surface, since the wafer has to be exposed to air while pumping the sample chamber (1-2 hours). The thickness of the native oxide layer is approximately 1~nm based on studies in Ref. \cite{wafer_oxide}. By assuming different thicknesses of the thin native oxide layer, the reflectance of the reference sample at different angles of incidence can be calculated based on Snell's law and Fresnel's equation; see more detailed discussions in section 5.3. Figure \ref{maxdiff} shows the calculated maximum percentage difference of the reflectance (over the full range of incident angles) between cases with and without the oxide layer for wavelengths from 120~nm to 250~nm. Oxide layer thicknesses of 1~nm and 2~nm were calculated and are presented as red and blue lines, respectively. At short wavelengths, the effect of the oxide layer on reflectance is much larger than that at longer wavelengths, and the thicker native oxide layer causes larger shifts in reflectance than those that occur in the case of no oxidation. Ratios of the measured reflectance and the calculated reflectance of the reference sample (assuming no oxide layer on the wafer) versus the angle of incidence are shown in Figure \ref{ratio} for ten selected wavelengths. The results of s-polarization (Figure \ref{ratio} (a)) and p-polarization (Figure \ref{ratio} (b)) are compared separately. For wavelengths of 128~nm and 150~nm, larger discrepancies are observed since the reflectance is more sensitive to the thickness of the native oxide layer on the reference sample surface. Thus, in uncertainty estimation, curves of 128~nm and 150~nm are excluded. For other wavelengths, the calculated results agree with the measurements within 8\% (rel.) for both s-polarization and p-polarization, and we take this number as the systematic error induced by the specular reflectance optical system. Uncertainties from other factors are negligible.

The absolute systematic uncertainty of the TIS is 20\%, as stated in its user manual. This uncertainty arises from the unavailability of calibrated commercial samples. Even though a micro-roughened ceramic silicon carbide (SiC) sample, which shows an excellent long-term stability against VUV laser radiation, is measured by a Perkin-Elmer spectrophotometer with an installed integral sphere to determine the total amount of diffuse reflectance, the environmental conditions between the two setups might be different. The relative uncertainty of the TIS is 5\%, also quoted from the user manual, which means that the diffuse reflectance of different samples measured by the TIS can be compared with a relatively high accuracy.

\section{Results}
\subsection{Specular reflectance}

\begin{figure}
    \centering
    \subfigure{
			\centering
			\includegraphics[width=3in]{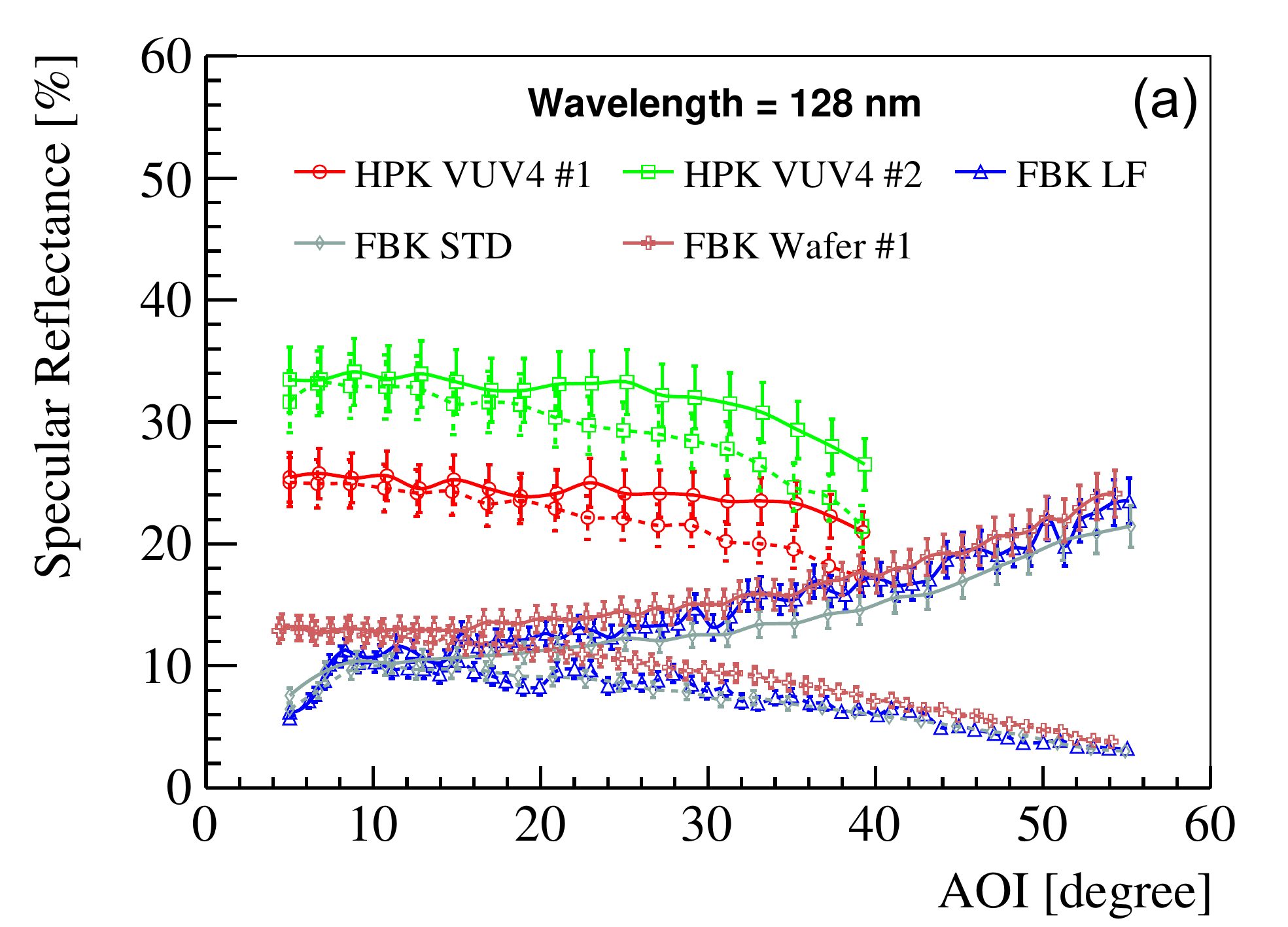}
			\label{group_128nm}
	}
	\subfigure{
			\centering
			\includegraphics[width=3in]{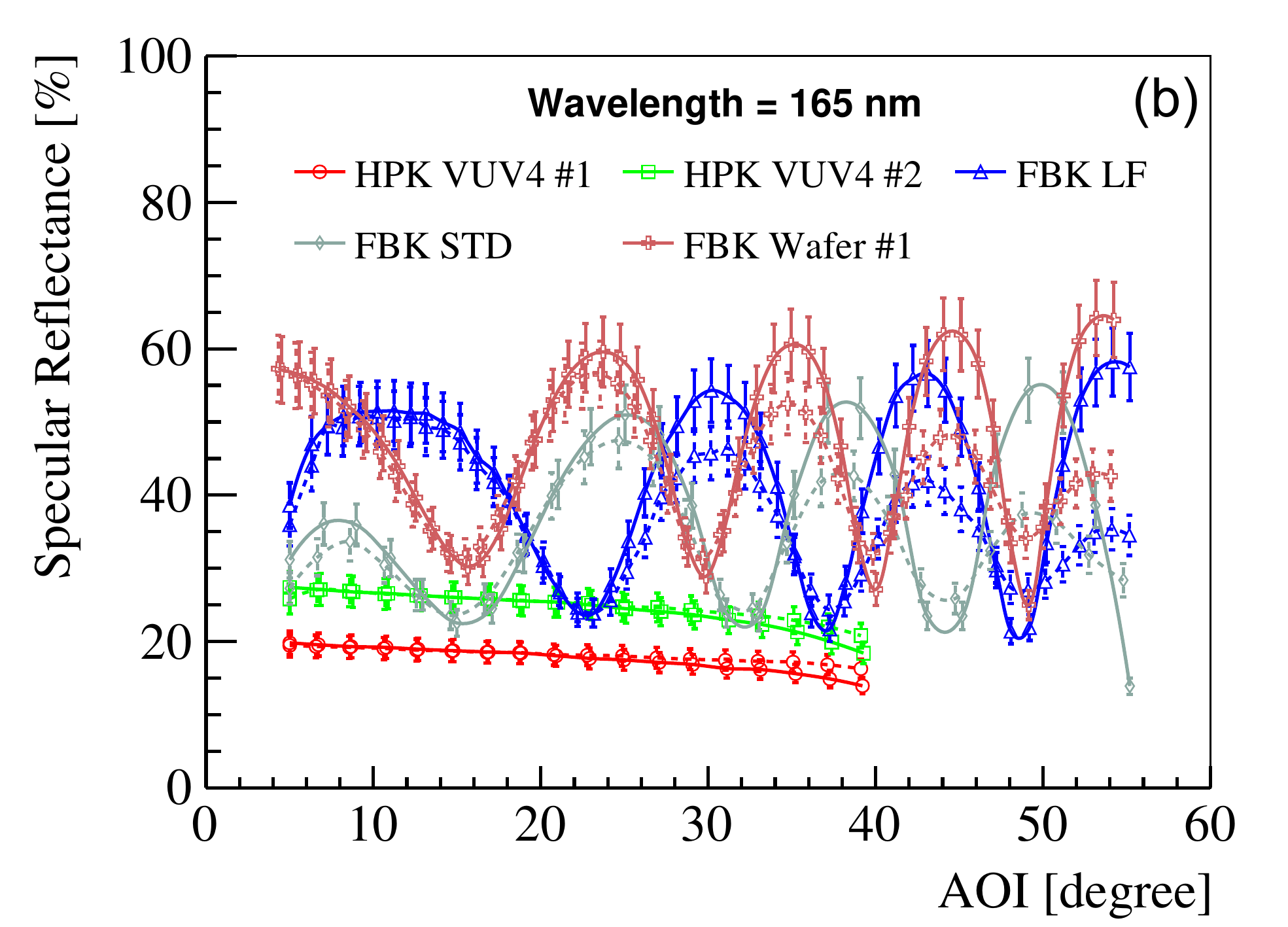}
			\label{group_165nm}
	}
	\vspace{.1in}
    \subfigure{
			\centering
			\includegraphics[width=3in]{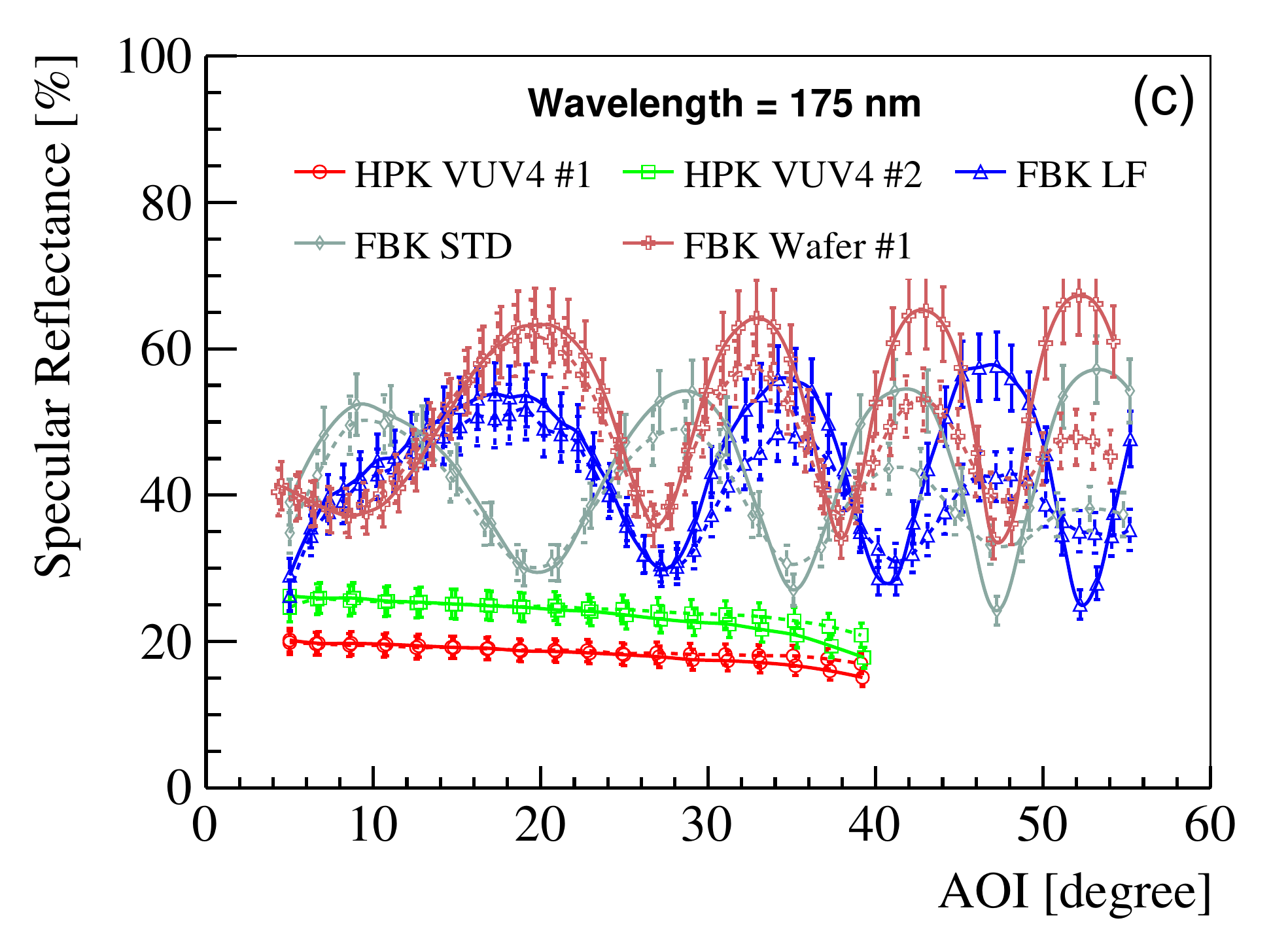}
			\label{group_175nm}
	}
    \subfigure{
			\centering
			\includegraphics[width=3in]{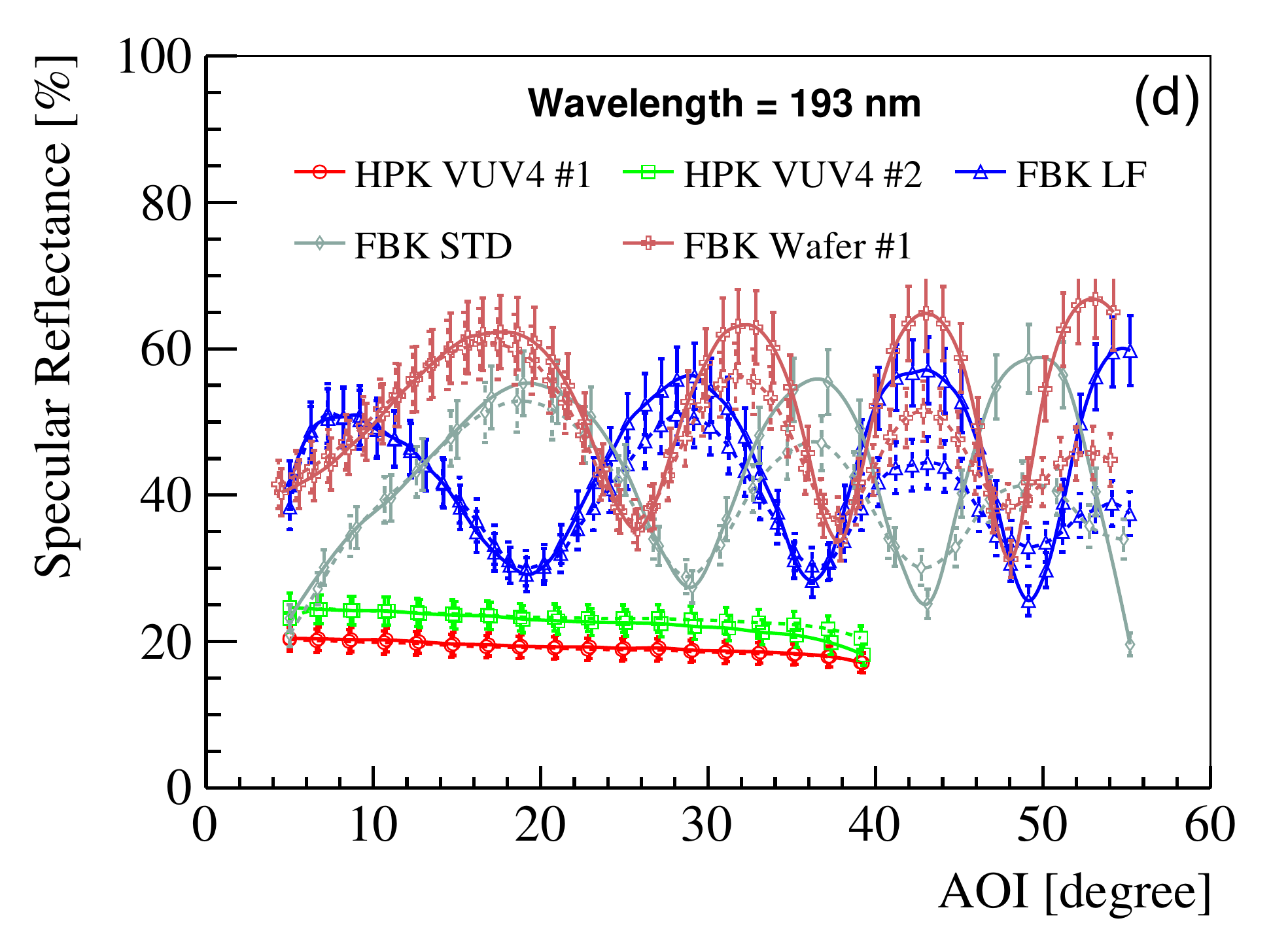}
			\label{group_193nm}
	}
        \caption{Specular reflectance as a function of AOI for five samples measured at four wavelengths: 128~nm, 165~nm, 175~nm and 193~nm. Solid lines represent light beams with s-polarization, while dashed lines indicate p-polarization.}    \label{groupall}
\end{figure}

The angle dependence of the specular reflectance is measured by the specular reflection optical system for the samples listed in Table \ref{tab:1}. Nine wavelengths are used in this measurement, covering the range from 128~nm to 200~nm. The results of four typical wavelengths are selected and shown in Figure \ref{groupall}, in which 128~nm represents the central wavelength of scintillation light emitted from liquid argon \cite{lar_scint}, 175~nm is the peak of the LXe emission spectrum \cite{lxe_scint}, and 193~nm is used in measuring the diffuse reflectance. The wavelength is labeled on each plot, and samples are indicated by different colors. The specular reflectance of five samples (FBK-Si-Wafer \#2 is not shown) is measured with s-polarization (represented as solid lines) and p-polarization (represented as dashed lines) light separately. In general, the two FBK SiPM samples reflect more light at long wavelengths than the HPK SiPMs. However at 128~nm, the trend is opposite for AOI less than 40 degree. The specular reflectance of the two VUV4 devices is found to decrease with the angle of incidence. Data above 40 degree, not shown in plots, are not accurate for two HPK SiPMs due to the shadowing effect of the sample holder. The sample holder is 1.5~mm higher than the sample surface, so part of the light is blocked at large incident angles. However, no such issue exists for the three FBK samples, because they have much larger dimensions. The sample of VUV4 \#2 with the larger pixel size shows higher specular reflectance than that of VUV4 \#1 due to its larger filling factor. The oscillation is not observed for HPK SiPMs but can clearly be seen in all three FBK samples at three longer wavelengths. This result is caused by the interference of incident light in the thin SiO$_2$ layer deposited on the surface of the FBK samples. The thickness of the thin SiO$_2$ layer is approximately 1.5~$\mu$m, as measured by FBK \cite{thick_sio2}. The FBK-Si-Wafer sample has higher specular reflectance than that of the other two FBK SiPMs, because the microstructures on the surface of SiPMs, such as traces, quenching resistors, etc., can reduce the specular reflectance. The FBK-VUV-HD1-LF and FBK-VUV-HD1-STD samples have almost the same reflectance, since they share similar profiles and surface structures. The shift of the oscillation phase between the two FBK SiPMs is introduced by the difference in the thickness of the SiO$_2$ layers on their surfaces. 

\begin{figure}
    \centering
    \subfigure{
			\centering
			\includegraphics[width=3in]{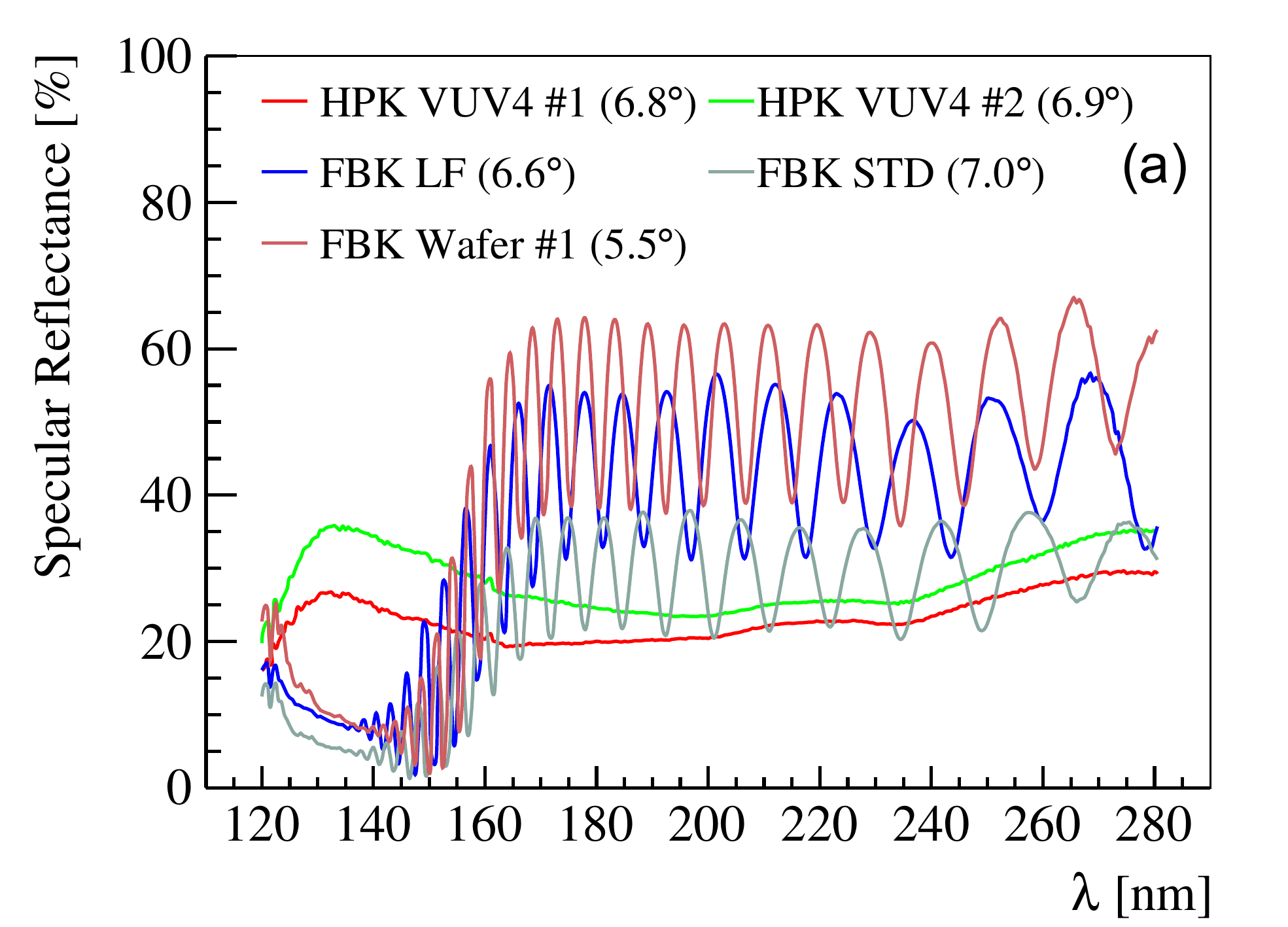}
			\label{group_7deg}
	}
	\subfigure{
			\centering
			\includegraphics[width=3in]{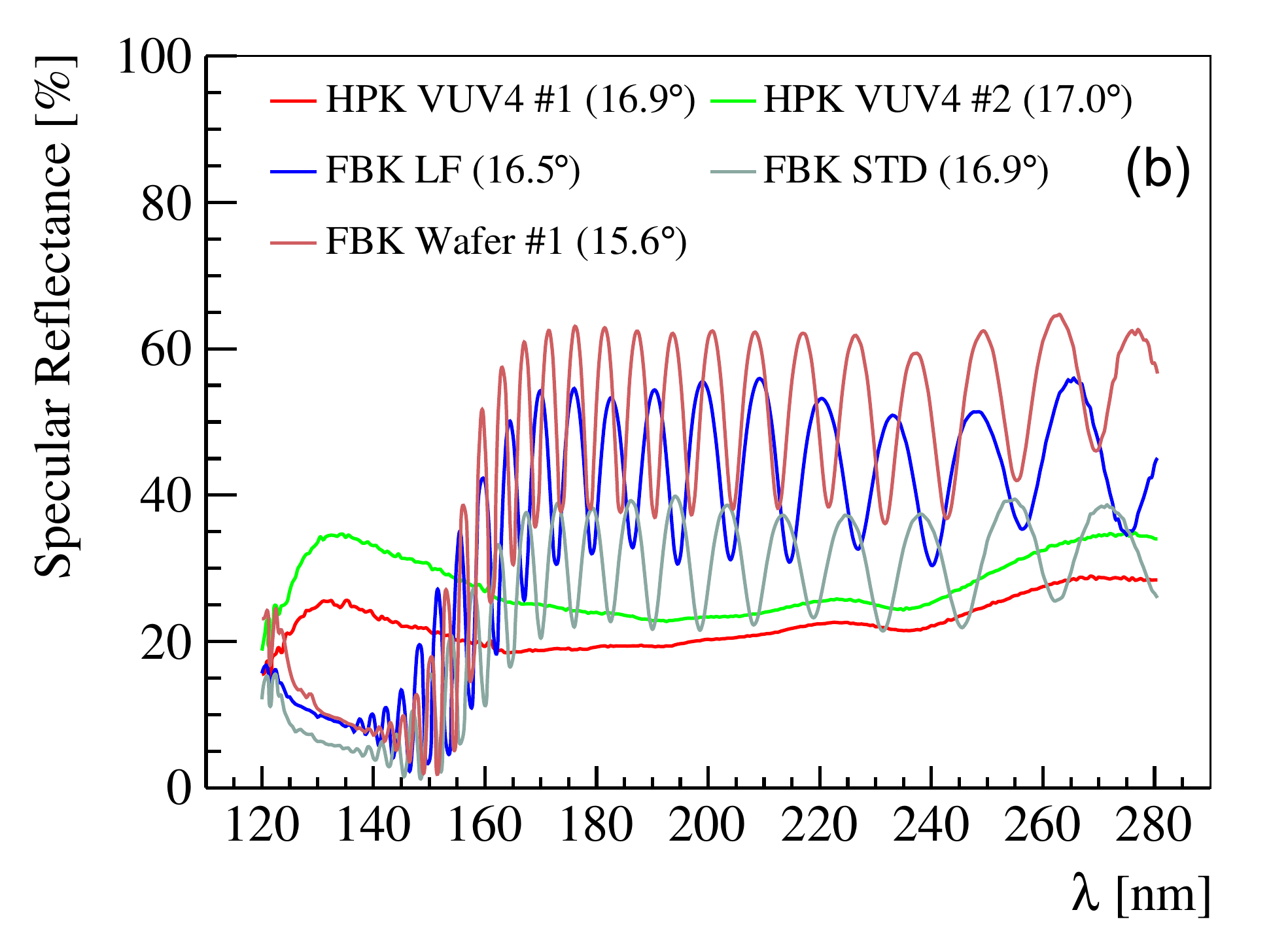}
			\label{group_17deg}
	}
    \subfigure{
			\centering
			\includegraphics[width=3in]{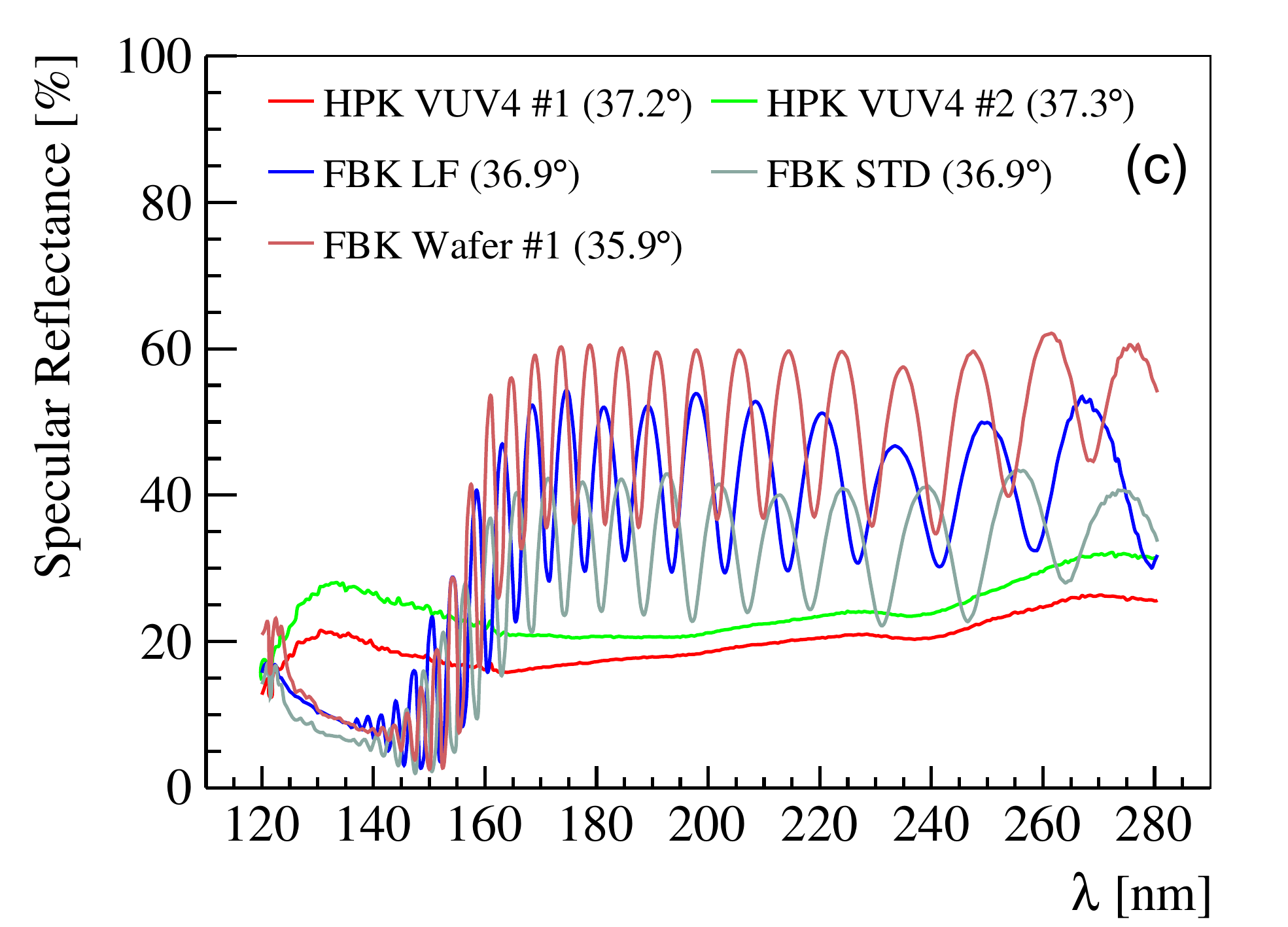}
			\label{group_37deg}
	}
    \subfigure{
			\centering
			\includegraphics[width=3in]{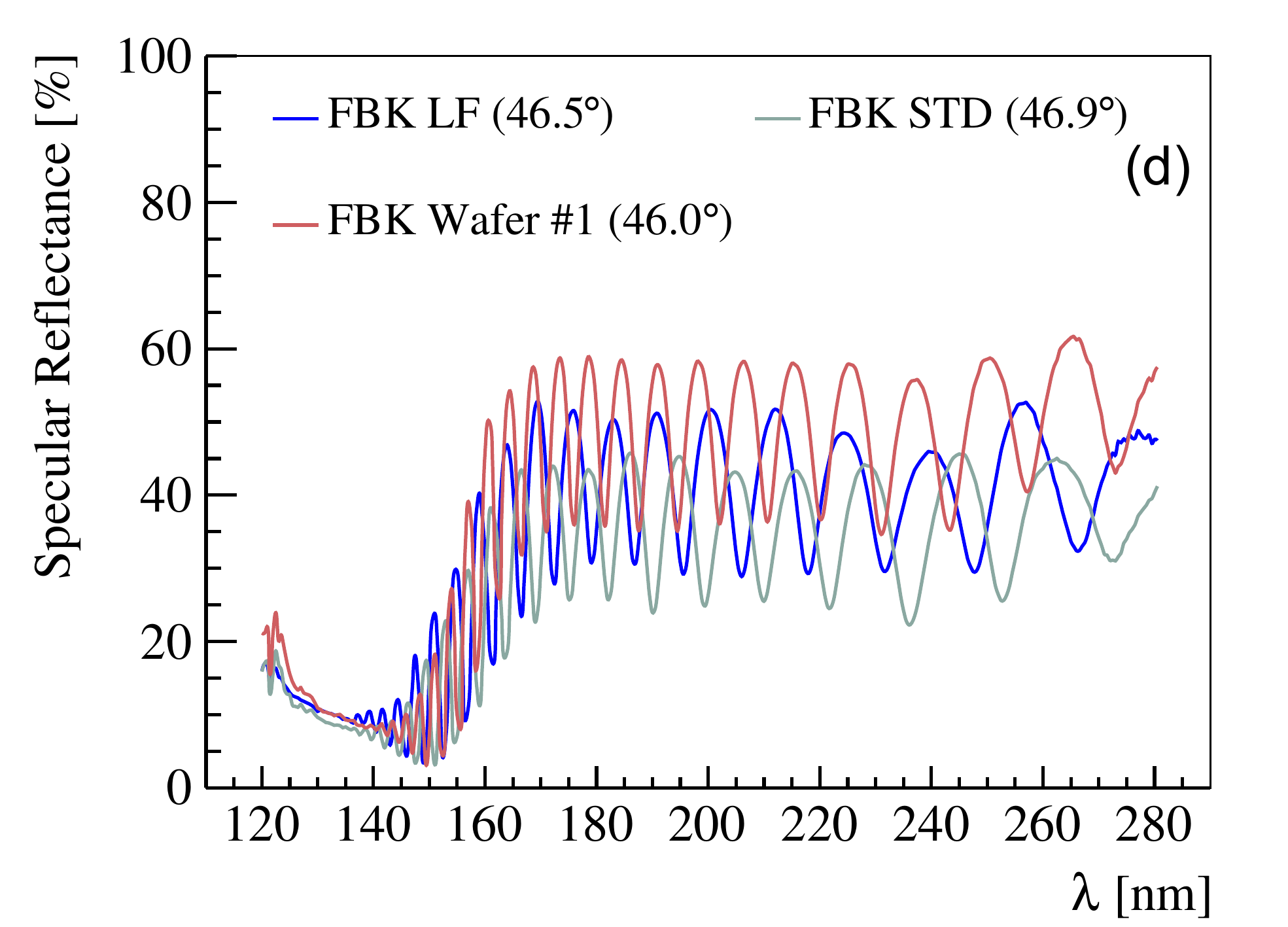}
			\label{group_57deg}
	}
        \caption{Specular reflectance as a function of wavelengths for five samples measured at different incident angles, as indicated in the plots.}
    \label{groupall_deg}
\end{figure}

Figure \ref{groupall_deg} presents the specular reflectance versus different wavelengths for samples measured at different incident angles. The non-polarized light beam is used in this measurement, and errors (rel. 8\%) are omitted for clarity. The wavelength covers a range from 120~nm to 280~nm. Different samples are marked with different colors in the plots, together with their incident angles, which are indicated in brackets. Data at AOI of $\sim$46 degree are not drawn for the HPK SiPMs, due to the aforementioned shadowing effect. Similar to Figure \ref{groupall}, the specular reflectance of the FBK samples oscillates with the wavelengths due to the interference induced by the thin SiO$_2$ layer. No oscillations are observed for the two HPK SiPMs. VUV4 \#2 has larger specular reflectance in the range of measured wavelengths because of its larger filling factor. The specular reflectance of the two FBK SiPMs is slightly lower than that of the silicon wafer, as expected. The low reflectance of FBK-VUV-HD1-STD at selected incident angles compared with that of FBK-VUV-HD1-LF is caused by the different oscillation phases determined by the different thicknesses of the SiO$_2$ layer. 

\subsection{Diffuse reflectance}
The diffuse reflectance of the five samples was measured by the TIS in vacuum. The TIS scanned each sample to obtain the diffuse reflectance at different positions. The average diffuse reflectance within the inscribed circle of each sample is presented in Table {\ref{tab:2}}, where contributions from the specular reflected light, left from the entrance port on the TIS, are not included. The FBK silicon wafer shows a negligible amount of the diffuse component, due to its mirror-like surface. For SiPMs, a relatively large fraction of diffuse reflections are observed at a level of 10\%. To compare the two SiPMs of HPK, the SiPM with the larger filling factor has lower diffusion, since the diffusion is mainly caused by the microstructure on the surface of the SiPM. For the two FBK SiPMs, the diffuse reflectance is similar to that of the HPK SiPMs. In addition to the amount of the absolute diffuse component, the angle response of diffuse reflections is also very interesting and an important input for detector simulation. The TIS does not have the ability to study this feature; instead, it will be studied in the future based on other ongoing setups.

\begin{table}
	\centering
	\caption{\label{tab:2} Results of diffuse reflectance for 2 FBK and 2 HPK SiPMs. For comparison, a silicon wafer from FBK has been measured as a reference.}
	\smallskip
	\begin{tabular}{|c|c|}
		\hline
		Sample Name& Diffuse reflectance\\
		\hline	
		FBK-Si-Wafer \#1    &($0.10\pm0.02$)\% \\
		FBK-VUV-HD1-LF      &($10.0\pm2.0$)\% \\
		FBK-VUV-HD1-STD    &($13.3\pm2.7$)\%	\\
		Hamamatsu-VUV4 \#1	  &($17.5\pm3.5$)\%	\\
		Hamamatsu-VUV4 \#2    &($10.0\pm2.0$)\%	\\
		\hline
	\end{tabular}
\end{table}

\subsection{Optical properties of the SiO$_2$ film}
The optical properties of the antireflective coating deposited on the SiPM surface possibly depend on the technologies used to produce the film. For the FBK SiPMs discussed in this work, a SiO$_2$ layer with the thickness of approximately 1.5~$\mu$m was deposited onto the silicon surfaces. The optical properties of the SiO$_2$ film, such as the refractive index ($n$) and extinction coefficient ($k$), can be extracted by analyzing the reflectance data of the FBK-Si-Wafer sample, because the SiO$_2$ films on the FBK-Si-Wafer and FBK SiPM were produced based on the same technologies, and almost no diffuse reflections occur on the FBK-Si-Wafer, which makes it easier to extract its $n$ and $k$. For a two-media system, the reflection coefficient (the ratio of the electric field amplitudes of the incident light and reflected light) of light with s-polarization and p-polarization can be calculated by Fresnel's equation:
\begin{equation}
\begin{array}{ll}
	{r_{s}=\frac{\tilde{n}_{0} \cos \theta_{0}-\tilde{n}_{1} \cos \theta_{1}}{\tilde{n}_{0} \cos \theta_{0}+\tilde{n}_{1} \cos \theta_{1}}  } & {~~~~~~r_{p}=\frac{\tilde{n}_{1} \cos \theta_{0}-\tilde{n}_{0} \cos \theta_{1}}{\tilde{n}_{0} \cos \theta_{0}+\tilde{n}_{1} \cos \theta_{1}},} 
\end{array}
\end{equation}
in which $\Tilde{n}_{0}$ and $\Tilde{n}_{1}$ are complex indices of refraction of the first medium and the second medium, respectively, which are functions of the wavelength ($\lambda$) and can be expressed in terms of $n$ and $k$:
\begin{equation}
		\Tilde{n}=n(\lambda)+i k(\lambda),
\end{equation}
$\theta_{0}$ denotes the incident angle of the light beam, and $\theta_{1}$ represents the refractive angle. The relation between $\theta_{0}$ and $\theta_{1}$ is determined by Snell's law:
\begin{equation}
\begin{array}{ll}
	\Tilde{n}_{0} \sin \left(\theta_{0}\right)=\Tilde{n}_{1} \sin \left(\theta_{1}\right),
\end{array}
\end{equation}
For the FBK-Si-Wafer sample, the SiO$_2$ film can be taken as a membrane; in this case, multiple reflections in the film will occur, and reflected light beams will interfere with each other. The total reflectance of light with s-polarization ($R_s$) and p-polarization ($R_p$) should be the superposition of all reflections. Based on equations 5.1 and 5.2, $R_s$ and $R_p$ can be easily derived as 

\begin{equation}
	R_{s} = \left|r_{s}\right|^{2} = |\frac{r_{s01}+e^{2 i \delta} r_{s12}}{1-e^{2 i \delta} r_{s01} r_{s12}}|^{2}, 
\end{equation}
\begin{equation}
	R_{p} = \left|r_{p}\right|^{2} = |\frac{r_{p01}+e^{2 i \delta} r_{p12}}{1-e^{2 i \delta} r_{p01} r_{p12}}|^{2}, 
\end{equation}
$r_{s01}$ ($r_{p01}$) and $r_{s12}$ ($r_{p12}$) represent the reflection coefficients of light with s-polarization (p-polarization) from vacuum to SiO$_2$ and from SiO$_2$ to silicon, respectively. $\delta$ is the phase difference between two adjacent light beams, determined by 
\begin{equation}
\begin{array}{ll}
\delta=\frac{2 \pi d_{1}}{\lambda} \tilde{n}_{1} \cos \theta_{1},
\end{array}
\end{equation}
$d_{1}$ denotes the thickness of the SiO$_2$ film, $\lambda$ is the wavelength of the incident light, and $\tilde{n}_{1}$ and $\theta_{1}$ are the complex refractive index and refractive angle in SiO$_2$, respectively. For non-polarized light, the reflectance is an average of the reflectance of s-polarization and p-polarization.
\begin{equation}
\begin{array}{ll}
	R=\frac{1}{2}\left(R_{S}+R_{P}\right) 
\end{array}
\end{equation}

\begin{figure}
    \centering
    \subfigure{
			\centering
			\includegraphics[width=3in]{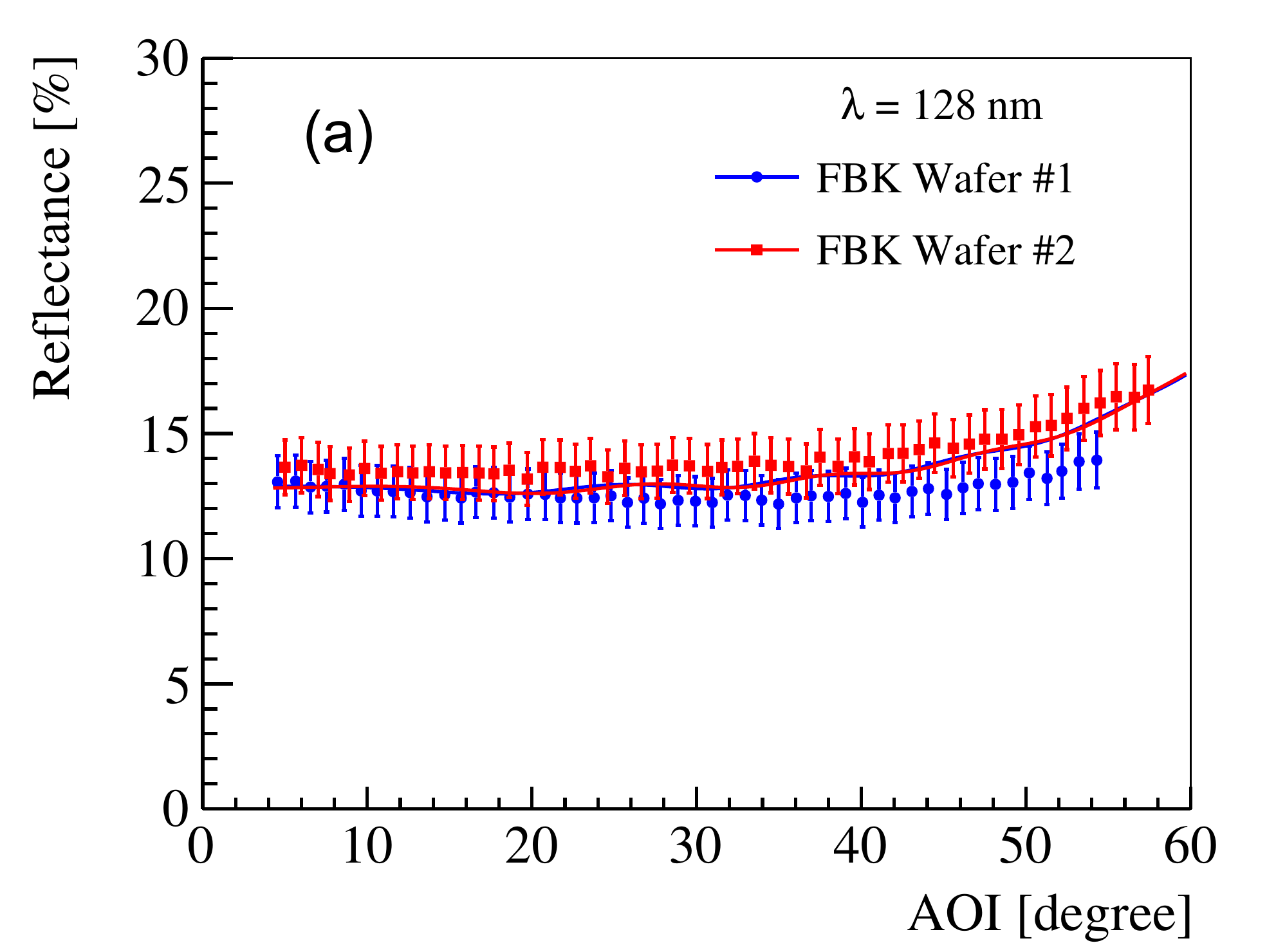}
			\label{fit_128nm}
	}
	\subfigure{
			\centering
			\includegraphics[width=3in]{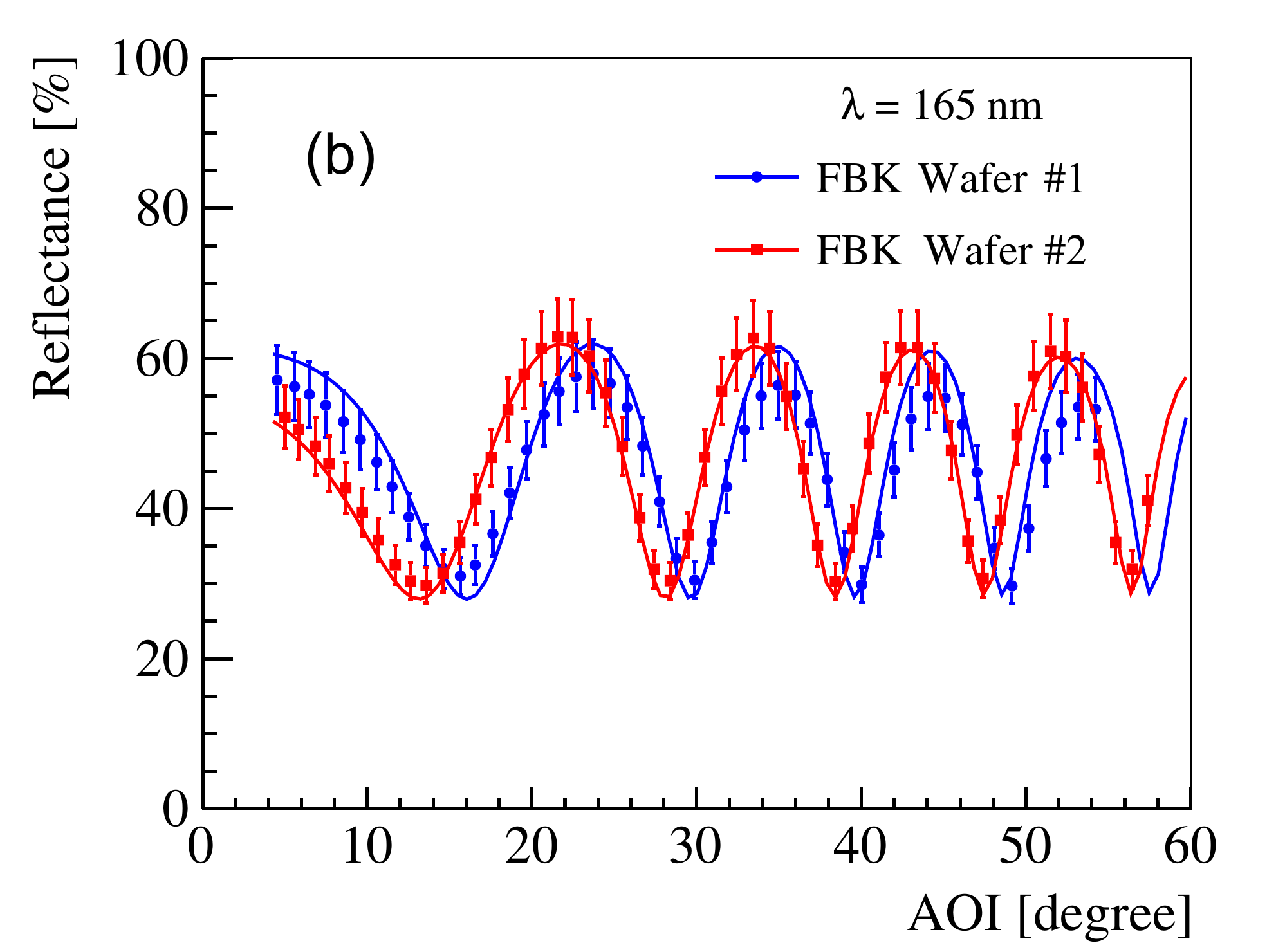}
			\label{fit_165nm.png}
	}
    \subfigure{
			\centering
			\includegraphics[width=3in]{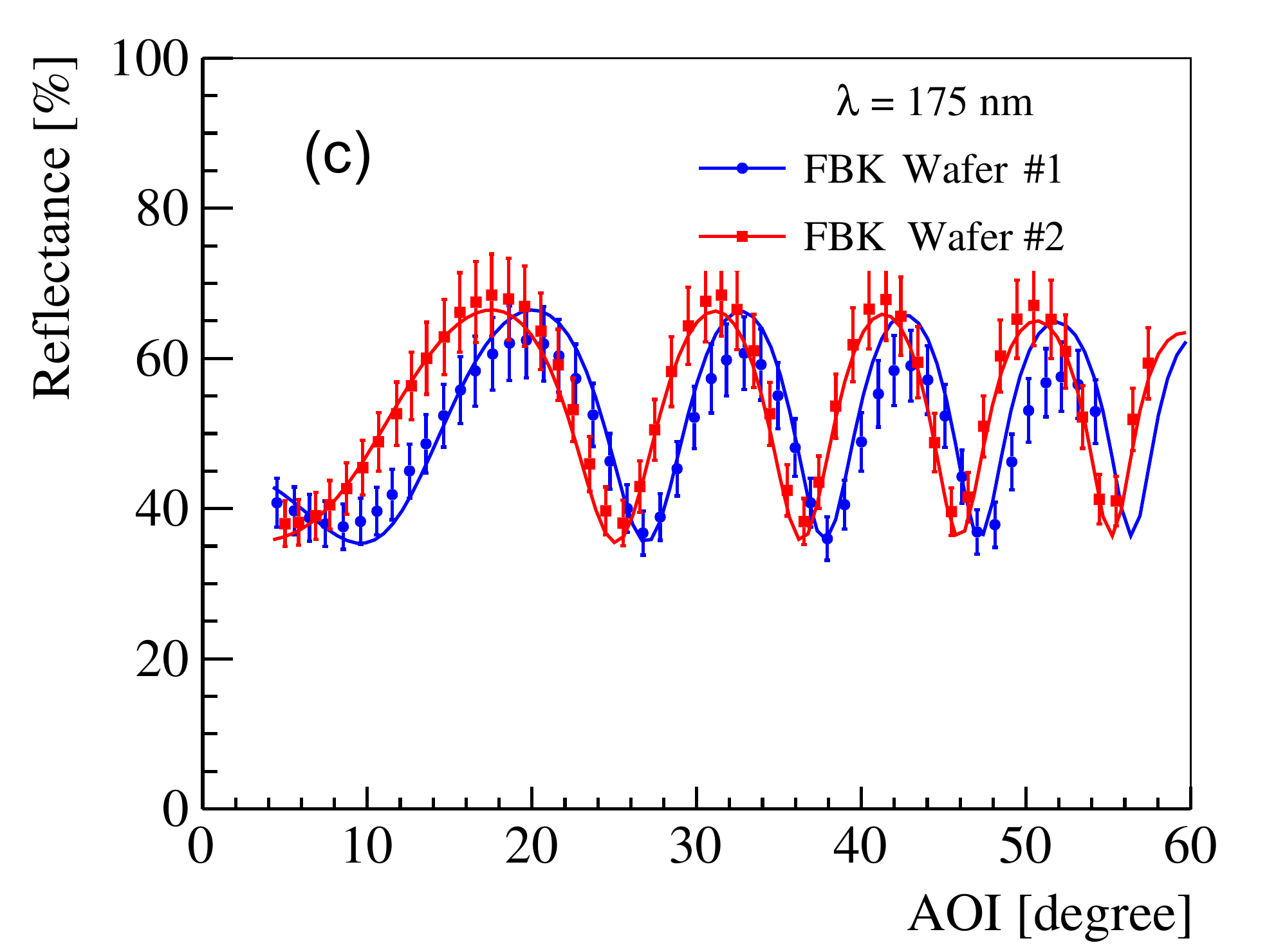}
			\label{fit_175nm}
	}
    \subfigure{
			\centering
			\includegraphics[width=3in]{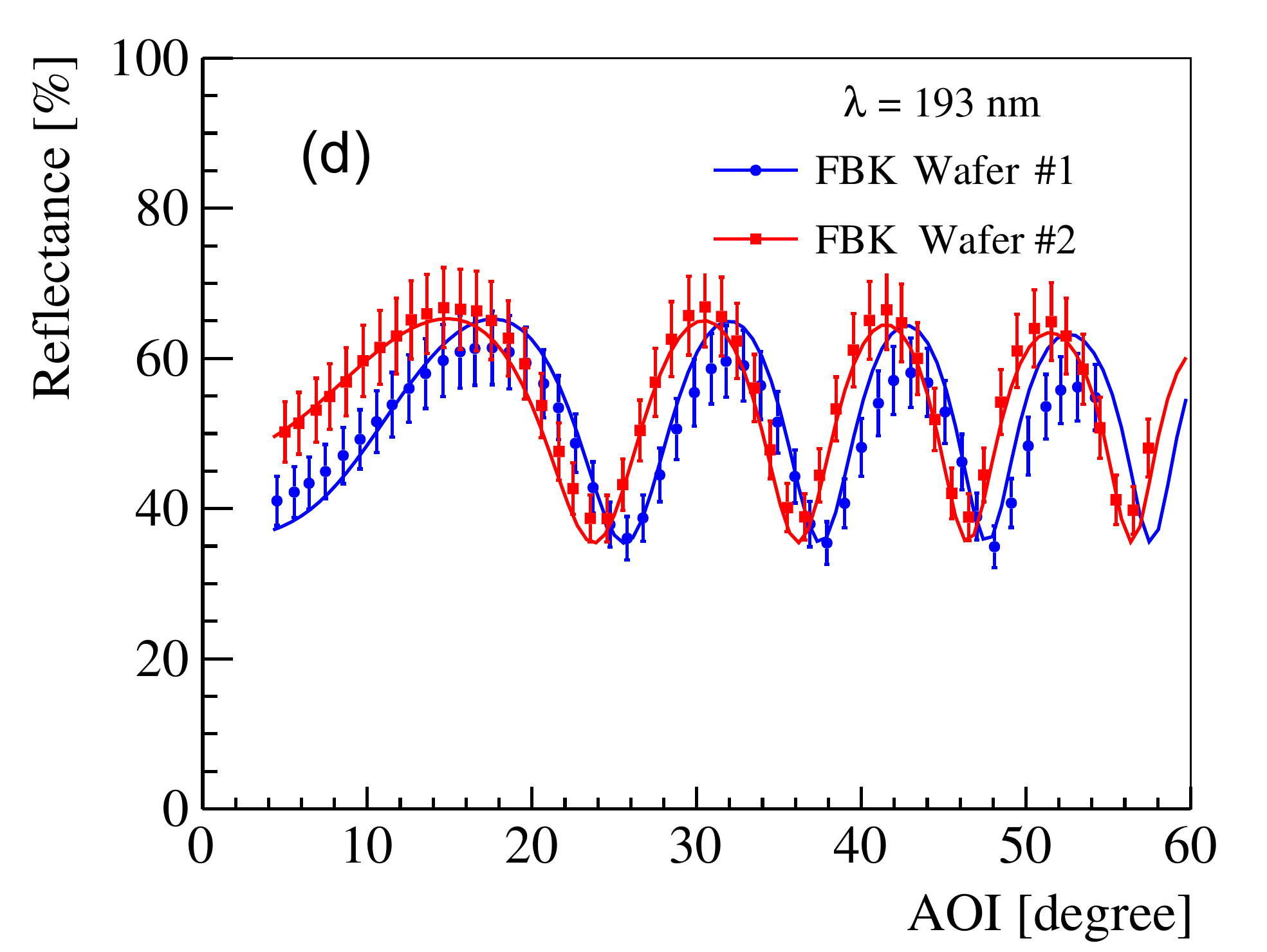}
			\label{fit_193}
	}
        \caption{Reflectance as a function of AOI measured at four wavelengths for the two FBK silicon wafer samples. The fitted curves are shown as solid lines.}
    \label{fitall}
\end{figure}

For FBK-Si-Wafer \#1 and FBK-Si-Wafer \#2, the reflectance versus the angle of incidence was measured at 9 different wavelengths, in which the FBK-Si-Wafer \#1 was measured by light with s-polarization and p-polarization separately and the \#2 sample was measured with non-polarized light. A customized fitting program is developed based on TMinuit \cite{tminuit} to simultaneously fit the reflectance data of FBK-Si-Wafer \#1 and FBK-Si-Wafer \#2 by using equation 5.7. Non-polarized reflectance data are used for both samples during the fitting. The $n$ and $k$ of silicon are from Ref.\cite{si_nk} and fixed in the fit. The $n$ and $k$ of SiO$_2$ and the two thicknesses of the two samples are the four floating parameters. As examples, the fitted results at four wavelengths are shown in Figure \ref{fitall}. The fitted curves well reproduce the measurements. The amplitudes of reflectance of the two samples are identical within the uncertainty, and the phase differences in plots (b), (c), and (d) are caused by the different thicknesses of oxide layers on surfaces of the two samples. From the fitting, the average oxide-layer thicknesses of the \#1 and \#2 samples are found to be 1.519 $\pm$ 0.008 $\mu$m and 1.512 $\pm$ 0.008 $\mu$m, respectively. The oxide-layer thickness of sample \#1 is also measured by an ellipsometer and determined to be 1.527 $\pm$ 0.004 $\mu$m, which agrees with the value determined in reflectance measurements. The maximum difference in oxide-layer thickness of sample \#1 between the two measurements (rel. 1.5\%) is taken as an uncertainty and added to the errors of the fitted $n$ and $k$ of SiO$_2$, which are shown in Figure \ref{n,k}. The refractive index of the SiO$_2$ film from our measurements is slightly lower than the numbers measured in Ref. \cite{sio2_nk}, as indicated by the black line, which might be a result of the different thickness of the film and manufacturing technologies used to make the film. For the extinction coefficient of SiO$_2$, our data do not have good constraints at longer wavelengths, due to the very weak absorption in the SiO$_2$ film. Moreover, the fitted values at short wavelengths do not match those in Ref.\cite{sio2_nk}, as shown by the black line (see Figure \ref{n,k}), this may be caused by the aforementioned reasons with that of the refractive index. The $n$ and $k$ of SiO$_2$ are calculated by fixing the thickness of the SiO$_2$ film of the two samples to above-average values in the fit.

\begin{figure}
    \centering
    \subfigure{
		\begin{minipage}[b]{0.45\textwidth}
			\centering
			\includegraphics[width=3in]{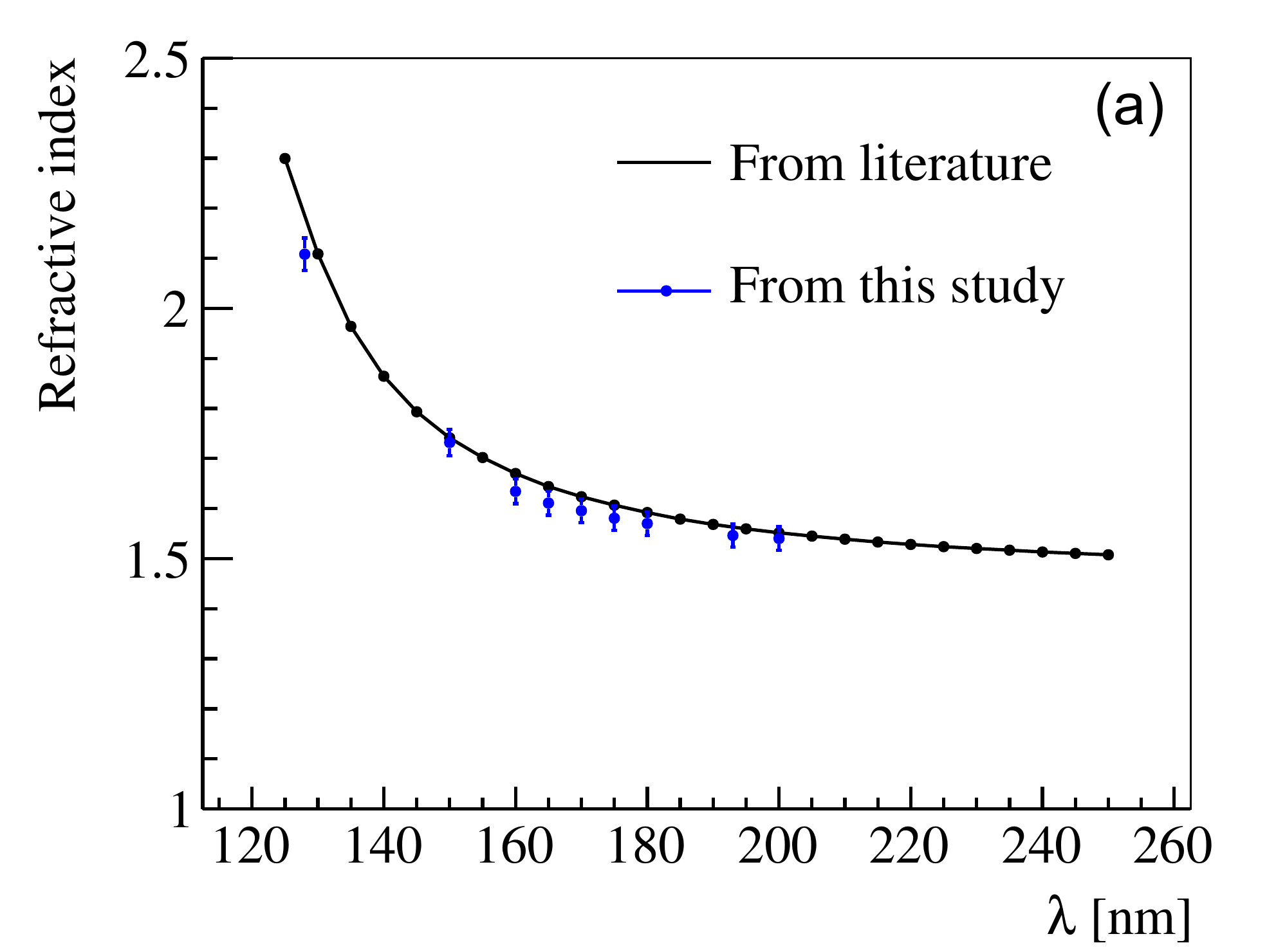}
			\label{sio2_n}
		\end{minipage}
	}
	\subfigure{
			\centering
			\includegraphics[width=3in]{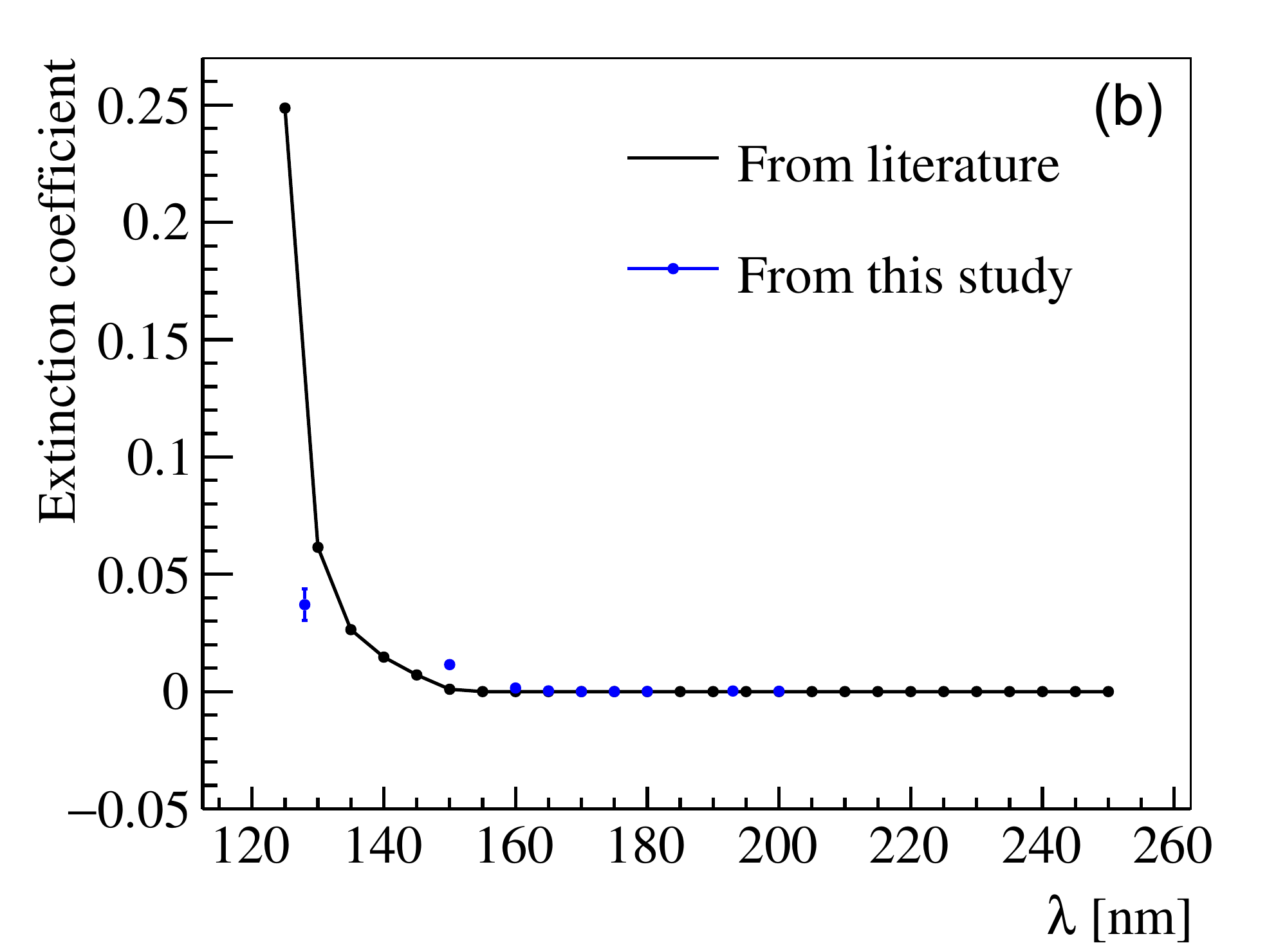}
			\label{sio2_k}
	}
        \caption{Fitted refractive index (a) and extinction coefficient (b) of $SiO_2$ film, compared to results from the literature \cite{sio2_nk} (black lines).}
    \label{n,k}
\end{figure}
\subsection{Prediction of reflectance in LXe}
\begin{figure}
\centering
\includegraphics[width=3in]{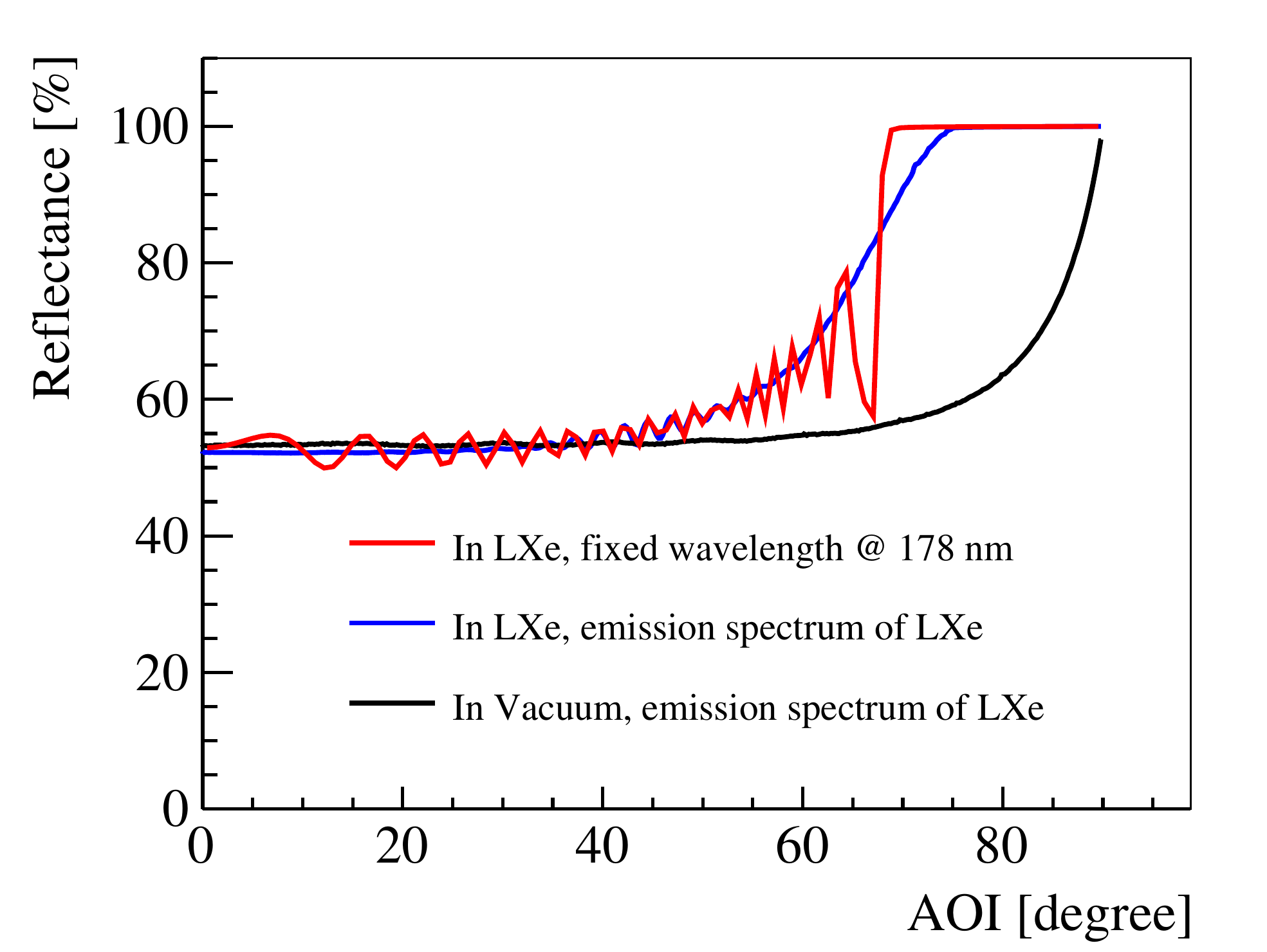}
\caption{Predicted reflectance as a function of AOI for the FBK silicon wafer. Red line: the reflectance in LXe for incident light with a fixed wavelength of 178~nm; Blue line: the reflectance in LXe for incident light with wavelengths that follow the distribution of the LXe emission spectrum. Black line: similar with the blue line, but calculated in vacuum.} 
\label{lxe}
\end{figure}

In the nEXO TPC, the SiPM array will be operated in LXe, reflectance of SiPMs in LXe is desired. In principle, the reflectance of samples in LXe can be predicted based on the known composition and thickness and their refractive indices and extinction coefficients, in particular for samples with a mirror-like surface. For SiPMs, this prediction becomes difficult due to the complex layout and materials of the microstructure on its surface, but for the specular reflection component, it should be possible. In this work, we calculate the reflectance of the FBK silicon wafer in LXe based on the $n$ and $k$ of SiO$_2$ film discussed in the previous section. The results are shown in Figure \ref{lxe}. The thickness of the SiO$_2$ layer on top of the silicon wafer is assumed to be 1.5 $\mu$m. Similar to that in a vacuum, the oscillation structure caused by interference can be observed in LXe for incident light with a fixed wavelength, shown as red curve in the figure. However, the amplitude of the oscillation is significantly suppressed in LXe. After taking the emission spectrum of liquid xenon (central wavelength is 175~nm; FWHM is 10~nm) \cite{lxe_scint} into account, the oscillation structure disappears both in vacuum and liquid xenon, shown as the black curve and blue curve, respectively, because the effect of the interference is canceled out by the wavelength variation of the incident light. The critical angle in liquid xenon becomes smaller than that in a vacuum; hence, total reflection can clearly be seen in liquid xenon. The calculated reflectance of the FBK-Si-Wafer in liquid xenon is $(52.2\pm1.6)\%$ at the incident angle of 15 degree, which is consistent with the number of $(50.8\pm2.3)\%$ measured at the same incident angle by the LXe-based setup in nEXO \cite{lixo}. More comparisons at different incident angles will be performed in the near future. The specular reflectance of the measured FBK SiPMs in LXe can be roughly estimated by applying the same scale factor from vacuum to LXe, obtained from the FBK silicon wafer. For Hamamatsu SiPMs, their reflectance in liquid xenon has to be measured by LXe-based setups, since we do not have any information on the ARC. 


%

\section{Conclusions}

We measured the specular and diffuse reflectance in a vacuum for the two FBK SiPMs (FBK-VUV-HD1-LF and FBK-VUV-HD1-STD) and two HPK SiPMs (VUV4 with a pixel size of 50~$\mu$m and 75~$\mu$m). The results show that SiPMs reflect a large fraction of VUV light. Furthermore, SiPMs from FBK are more reflective than those from HPK (VUV4). The diffuse component of reflective light is also observed, which is caused by the microstructures on the SiPM surface. The $n$ and $k$ of the SiO$_2$ film on the FBK silicon wafer are extracted by analyzing its reflectance data, which is an important input for the design of ARCs. Finally, the reflectance of the FBK silicon wafer in LXe is predicted based on the new $n$ and $k$ of the SiO$_2$ film and can be used to verify the output of LXe-based reflectance setups.

\section*{Acknowledgment}
We like to thank Dr. Cunding Liu from Institute of Optics and Electronics, CAS and Prof. Bincheng Li from University of Electronic Science and Technology of China for their substantial assistance with the reflectance setups. We gratefully acknowledge support from CAS-IHEP Fund for PRC$\backslash$US Collaboration in HEP. Support for nEXO comes from the the Office of Nuclear Physics of the Department of Energy and NSF in the United States, from NSERC, CFI, FRQNT, NRC, and the McDonald Institute (CFREF) in Canada, from IBS in Korea, from RFBR (18-02-00550) in Russia, and from CAS and NSFC in China.  

P.~Lv is with Institute of High Energy Physics, Chinese Academy of Sciences, Beijing 100049, China.

G.F.~Cao is with Institute of High Energy Physics, Chinese Academy of Sciences, Beijing 100049, China, and also with University of Chinese Academy of Sciences, Beijing, China.

L.J.~Wen, W.H.~Wu, Z.~Ning, X.S.~Jiang, W.~Wei, X.L.~Sun, J.~Zhao and Y.Y.~Ding are with Institute of High Energy Physics, Chinese Academy of Sciences, Beijing 100049, China.

S.~Al Kharusi, T.~McElroy
% ...
Department of Physics, Yale University, New Haven, CT 06511, USA.

K.G.~Leach and C.R.~Natzke are with Department of Physics, Colorado School of Mines, Golden, CO 80401, USA.

D.S.~Leonard is with IBS Center for Underground Physics, Daejeon 34126, Korea.

O.~Njoya is with Department of Physics and Astronomy, Stony Brook University, SUNY, Stony Brook, NY 11794, USA.

G.~Visser is with Department of Physics and CEEM, Indiana University, Bloomington, IN 47405, USA.

J.-L.~Vuilleumier is with LHEP, Albert Einstein Center, University of Bern, Bern CH-3012, Switzerland.

L.~Yang is with Physics Department, University of California, San Diego, CA 92093, USA.

\ifCLASSOPTIONcaptionsoff
  \newpage
\fi

\vfill


\end{document}